\documentclass[aps,prd,reprint,preprintnumbers,showpacs,nofootinbib,superscript address,longbibliography]{revtex4-2}
\usepackage[utf8]{inputenc}
\usepackage{amsmath,amsfonts,amssymb,amsfonts}
\usepackage{fontawesome}
\usepackage{comment}
\usepackage{siunitx}
\usepackage{verbatim}
\usepackage[english]{babel}
\usepackage{adjustbox}
\usepackage{relsize} 
\usepackage{makecell}
\usepackage{pifont}
\usepackage{slashed}
\usepackage{scalerel,stackengine}
\usepackage{hhline}
\usepackage{mathtools}
\usepackage[dvipsnames]{xcolor}
\usepackage{tcolorbox}
\usepackage{pifont}
\usepackage{booktabs}
\usepackage{tabularx}
\usepackage{siunitx}
\usepackage{multirow}
\usepackage{graphicx}
\usepackage{xstring}
\usepackage{etoolbox}
\usepackage{notoccite}
\usepackage{natbib}
\usepackage{mathrsfs}
\usepackage{lineno}
\usepackage{tensor}
\usepackage{accents}
\usepackage{tikz}
\usepackage{tikz-feynman}
\usepackage{mdframed}
\usepackage{tocbasic}
\DeclareTOCStyleEntry[numwidth=25pt,linefill=\bfseries\TOCLineLeaderFill]{tocline}{section}
\DeclareTOCStyleEntry[entryformat=\textit,numwidth=10pt,linefill=\TOCLineLeaderFill]{tocline}{subsection}
\DeclareTOCStyleEntry[entryformat=\textit,numwidth=10pt,linefill=\TOCLineLeaderFill]{tocline}{subsubsection}
\allowdisplaybreaks

\def\nn{\nonumber}

\newcommand{\gsim}{\lower.7ex\hbox{$\;\stackrel{\textstyle>}{\sim}\;$}}
\newcommand{\lsim}{\lower.7ex\hbox{$\;\stackrel{\textstyle<}{\sim}\;$}}

\newcommand{\SampleConsistent}{\color{SeaGreen}\faCheck\color{black}}%
\newcommand{\SampleInconsistent}{\color{red}\faRemove\color{black}}%
\newcommand{\SampleEmpty}{\color{gray}\faCircleO\color{black}}%
\newcommand{\SampleMore}{\color{blue}\faPlus\color{black}}%

\newcommand{\Consistent}{\Large\color{SeaGreen}\faCheck\color{black}}%
\newcommand{\Inconsistent}{\Large\color{red}\faRemove\color{black}}%
\newcommand{\Empty}{\Large\color{gray}\faCircleO\color{black}}%
\newcommand{\More}{\Large\color{blue}\faPlus\color{black}}%

\newcommand{\be}{\begin{equation}}
\newcommand{\ee}{\end{equation}}
\newcommand{\bea}{\begin{equation}\begin{aligned}}
\newcommand{\eea}{\end{aligned}\end{equation}}

\newrobustcmd{\Dis}[1]{%
	\tensor{K}{#1}%
}
\newrobustcmd{\Con}[1]{%
	\tensor{\Gamma}{#1}%
}
\newrobustcmd{\MAGg}[1]{%
	\tensor{g}{#1}
}
\newrobustcmd{\G}[1]{%
	\tensor{\eta}{#1}
}
\newrobustcmd{\MAGA}[1]{%
  \tensor{A}{#1}
}
\newrobustcmd{\GeneralG}[2]{%
	\tensor{\vphantom{l^{l^2}}\smash{\stackrel{\raisebox{-3pt}{\scalebox{0.55}{$({#1})$}}}{\scalebox{0.99}{$\alpha$}}}}{#2}%
}
\newrobustcmd{\YukawaG}[2]{%
	\tensor{\vphantom{l^{l^2}}\smash{\stackrel{\raisebox{-3pt}{\scalebox{0.55}{$({#1})$}}}{\scalebox{0.99}{$\nu$}}}}{#2}%
}
\newrobustcmd{\ChiralG}[2]{%
	\tensor{\vphantom{l^{l^2}}\smash{\stackrel{\raisebox{-3pt}{\scalebox{0.55}{$({#1})$}}}{\scalebox{0.99}{$\chi$}}}}{#2}%
}
\newrobustcmd{\GRG}[2]{%
	\tensor{\vphantom{l^{l^2}}\smash{\stackrel{\raisebox{-3pt}{\scalebox{0.55}{$({#1})$}}}{\scalebox{0.99}{$\gamma$}}}}{#2}%
}
\newrobustcmd{\MAGG}[2]{%
	\tensor{\vphantom{l^{l^2}}\smash{\stackrel{\raisebox{-3pt}{\scalebox{0.55}{$({#1})$}}}{\scalebox{0.99}{$\kappa$}}}}{#2}%
}
\newrobustcmd{\OperatorO}[2]{%
	\tensor*{\mathcal{O}}{^{{#1}}_{(#2)}}
}
\newrobustcmd{\ChiralD}[1]{%
\tensor{\mathcal{D}}{#1}
}
\newrobustcmd{\MAGH}[1]{%
  \tensor*{h}{#1}
}
\newrobustcmd{\MAGEFTH}[1]{%
  \tensor*{H}{#1}
}
\newrobustcmd{\PD}[1]{%
	\tensor{\partial}{#1}%
}
\newrobustcmd{\CD}[1]{%
	\tensor{\nabla}{#1}%
}
\newrobustcmd{\Kronecker}[1]{%
	\tensor*{\delta}{#1}%
}
\newrobustcmd{\Gen}[1]{%
	\tensor{\xi}{#1}%
}
\newrobustcmd{\GenT}[1]{%
	\tensor*{\xi}{#1^{\text{T}}}%
}
\newrobustcmd{\GenTr}[1]{%
	\tensor*{\tilde{\xi}}{#1}%
}
\newrobustcmd{\MAGF}[1]{%
	\tensor{\mathcal{F}}{#1}
}
\newrobustcmd{\RD}[1]{%
	\tensor{\nabla}{#1}
}
\newrobustcmd{\MAGR}[1]{%
	\tensor{\mathcal{R}}{#1}
}
\newrobustcmd{\MAGEFTR}[1]{%
	\tensor{{R}}{#1}
}
\newrobustcmd{\MAGQ}[1]{%
	\tensor{\mathcal{Q}}{#1}
}
\newrobustcmd{\MAGT}[1]{%
	\tensor{\mathcal{T}}{#1}
}
\newrobustcmd{\Riem}[1]{%
	\tensor{\mathcal{R}}{#1}
}
\newrobustcmd{\TranslationGroup}{%
	\mathrm{T}(4)
}
\newrobustcmd{\GeneralLinearGroup}{%
	\mathrm{GL}(4,\mathbb{R})
}
\newrobustcmd{\RigidAffineGroup}{%
	\mathrm{A}(4,\mathbb{R})
}
\newrobustcmd{\LorentzGroup}{%
	\mathrm{SO}^+(3,1)
}
\newrobustcmd{\DiffeomorphismGroup}{%
	\mathrm{Diff}(4)
}
\newrobustcmd{\ConformalGroup}{%
	\mathrm{C}(3,1)
}

\newrobustcmd{\Model}[2]{%
	{$\tensor*{{\mathfrak{M}}}{_{#1}^{#2}}$}%
}
\newrobustcmd{\SubModel}[2]{%
	{$\tensor*{{\mathfrak{m}}}{_{#1}^{#2}}$}%
}
\newrobustcmd{\BadSquared}{%
	{\ding{55}${}^2$}
}
\newrobustcmd{\Bad}{%
	{\ding{55}}
}

\newrobustcmd{\UnknownMasslessParticle}{%
	{\big\{\tensor*{J}{^{P}_{\gamma}}\big\}}%
}
\newrobustcmd{\MasslessParticle}[2]{%
	{\tensor*{#1}{^{#2}_{\gamma}}}
}
\newrobustcmd{\MassiveParticle}[2]{%
	{\tensor*{#1}{^{#2}_{\text{m}}}}
}
\newrobustcmd{\MPl}{%
	{\mathrm{M}_{\text{Pl}}}
}

\usepackage{xspace}
\newrobustcmd{\PSALTer}{\textit{PSALTer}\xspace}

\tikzfeynmanset{
solid scalar/.style={draw=black, thick}
}
\usetikzlibrary{shapes.geometric, arrows, positioning}
\tikzstyle{input} = [rectangle, draw, fill=gray!20, minimum size=1cm, text centered]
\tikzstyle{inputT} = [rectangle, draw, fill=blue!20, minimum size=1cm, text centered]
\tikzstyle{hidden} = [circle, draw, fill=green!30, minimum size=1cm, text centered]
\tikzstyle{output} = [rectangle, draw, fill=white!30, minimum size=1cm, text centered]
\tikzstyle{spec} = [rectangle, draw, fill=red!30, minimum size=1cm, text centered]
\tikzstyle{arrow} = [thick,->,>=stealth]
\tikzstyle{arrow2} = [ultra thick, ->, >=latex, draw=black]
\tikzstyle{linec} = [ultra thick, -, >=latex, draw=black]
\tikzfeynmanset{compat=1.1.0}

\usepackage{hyperref}
\hypersetup{%
colorlinks = true,%
linkcolor = Blue,%
citecolor = Blue,%
filecolor = Blue,%
urlcolor = Blue%
}%
\usepackage[capitalize]{cleveref}
\AddToHook{cmd/appendix/before}{\crefalias{section}{appendix}}
\makeatletter
\renewcommand{\paragraph}{%
\@startsection{paragraph}{4}%
{\z@}{1.21ex \@plus 1ex \@minus .2ex}{0.9em}%
{\normalfont\normalsize\bfseries}%
}
\usepackage{titlesec}
\newrobustcmd{\pea}[1]{\emph{#1}\textbf{.\ \ \ ---}}
\titleformat{\paragraph}[runin]{\normalfont\normalsize\bfseries}{\emph\theparagraph}{1em}{\pea}
\titleformat{\section}[block]{\normalfont\bfseries\centering}{\MakeUppercase\thesection}{1em}{\MakeUppercase}
\makeatother

\makeatletter \def\switch@array{}\makeatother

\begin{document}

\title{Can metric-affine gravity be saved?}

\author{Will Barker}
\email{barker@fzu.cz}
\affiliation{Central European Institute for Cosmology and Fundamental Physics, Institute of Physics of the Czech Academy of Sciences, Na Slovance 1999/2, 182 00 Prague 8, Czechia}

\author{Carlo Marzo}
\email{carlo.marzo@kbfi.ee}
\affiliation{Laboratory for High Energy and Computational Physics, NICPB, R\"{a}vala 10, Tallinn 10143, Estonia}

\author{Alessandro Santoni}
\email{asantoni@uc.cl}
\affiliation{Institut f\"ur Theoretische Physik, Technische Universit\"at Wien, Wiedner Hauptstrasse 8--10, A-1040 Vienna, Austria}
\affiliation{Facultad de F\'isica, Pontificia Universidad Cat\'olica de Chile, Vicu\~{n}a Mackenna 4860, Santiago, Chile}

\begin{abstract}
Like general relativity, metric-affine gravity should be a viable effective quantum theory, otherwise it is a mathematical curiosity without physical application. Assuming a perturbative quantum field theory, the universal, flat limit of metric-affine gravity offers a good foundation for model-building only when symmetry constraints are themselves sufficient to get rid of ghosts and tachyons in the spectrum of propagating particle states, without requiring any further tuning of the couplings. Using this symmetry-first criterion, we find that for parity-preserving models with a totally symmetric distortion, only massless spin-one and spin-three modes are possible besides the graviton. Moreover, no viable models result from gauge symmetries generated by a scalar field.
\end{abstract}

\maketitle

\tableofcontents

\begin{figure}[htbp]
\includegraphics[width=\linewidth]{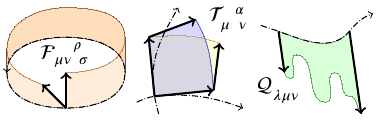}
\caption{\label{NonRiemannianSchematic} From~\cref{MAGFDef} curvature~$\MAGF{_{\mu\nu}^\rho_\sigma}$ rotates a vector after parallel transport around a closed loop. From~\cref{MAGTDef} torsion~$\MAGT{_\mu^\alpha_\nu}$ causes non-closure of parallelograms formed from (infinitesimally) parallel-transported vectors. From~\cref{MAGQDef} non-metricity~$\MAGQ{_{\lambda\mu\nu}}$ causes the evolution of vector length (norm) under parallel transport.}
\end{figure}

\section{Introduction}\label{Sec:Intro}

\paragraph*{Geometry vs physics} Metric-affine gravity (MAG) is a proposed model in which the metric~$\MAGg{_{\mu\nu}}$ and the connection~$\MAGA{_\mu^\rho_\nu}$ are treated as two independent fields~\cite{Hehl:1994ue}.
This leads to a richer spacetime geometry than allowed in general relativity (GR), characterised by curvature~$\MAGF{_{\mu\nu}^\rho_\sigma}$, torsion~$\MAGT{_\mu^\alpha_\nu}$ and non-metricity~$\MAGQ{_{\lambda\mu\nu}}$ -- respectively defined by the formulae
\begin{subequations}
\begin{align}
\MAGF{_{\mu\nu}^\rho_\sigma} &\equiv 2\left(\tensor{\partial}{_{[\mu}}\MAGA{_{\nu]}^\rho_\sigma}+\MAGA{_{[\mu|}^\rho_\alpha}\MAGA{_{|\nu]}^\alpha_\sigma}\right) \,, \label{MAGFDef} \\
\MAGT{_\mu^\alpha_\nu} &\equiv 2\MAGA{_{[\mu|}^\alpha_{|\nu]}} \,, \label{MAGTDef} \\ 
\MAGQ{_{\lambda\mu\nu}} &\equiv -\tensor{\partial}{_\lambda}\MAGg{_{\mu\nu}}+2\MAGA{_\lambda^\alpha_{(\mu|}}\MAGg{_{\alpha|\nu)}} \,. \label{MAGQDef}
\end{align}
\end{subequations}
The properties in~\cref{MAGFDef,MAGTDef,MAGQDef} have a popular geometric interpretation, which is illustrated in~\cref{NonRiemannianSchematic}.
GR is sometimes presented as an inherently \emph{geometric} theory, a viewpoint reinforced by its universal coupling to energy-momentum. In light of this, the picture in~\cref{NonRiemannianSchematic} seems to suggest that~$\MAGF{_{\mu\nu}^\rho_\sigma}$,~$\MAGT{_\mu^\alpha_\nu}$ and~$\MAGQ{_{\lambda\mu\nu}}$ constitute a more general and natural set of building blocks of gravity. 
As usually presented, however, this philosophy does not have any grounding in actual physics, the laws of which are completely insensitive to our geometric interpretation of the fields. Moreover, unlike in the case of GR, it does not constitute a particularly useful or constructive analogy. In sharp contrast to the simple Einstein--Hilbert theory, interactions described by the extended geometrical Lagrangian typically lack protection against detuning arising from quantum corrections.

\paragraph*{A theory of matter, not gravity} To better illustrate the problem, we perform the field reparameterisation
\begin{equation}\label{postriem}
\MAGA{_\mu^\rho_\nu} \equiv \Con{_\mu^\rho_\nu} +\Dis{_\mu^\rho_\nu}\,,
\end{equation}
where~$\Con{_\mu^\nu_\rho}\equiv\MAGg{^{\nu\lambda}}\big(\PD{_{(\mu}}\MAGg{_{\rho)\lambda}}-\frac{1}{2}\PD{_{\lambda}}\MAGg{_{\mu\rho}}\big)$ are the Christoffel symbols of the Levi--Civita connection from GR, and~$\Dis{_\mu^\rho_\nu}$ is an independent, rank-three, tensor field called the \emph{distortion}.
By plugging~\cref{postriem} into~\cref{MAGFDef}, we obtain (schematically, suppressing indices) the post-Riemannian decomposition~$\MAGF{}\sim\MAGR{}+\PD{}\Dis{}+\Dis{}^2$, where the textbook Riemannian curvature is
\begin{equation}\label{MAGRDef}
\MAGR{_{\mu\nu}^\rho_\sigma} \equiv 2\left(\tensor{\partial}{_{[\mu}}\Con{_{\nu]}^\rho_\sigma}+\Con{_{[\mu|}^\rho_\alpha}\Con{_{|\nu]}^\alpha_\sigma}\right).
\end{equation}
Similarly, by plugging~\cref{postriem} into~\cref{MAGTDef,MAGQDef}, we get formulae of the form~$\MAGT{}\sim\Dis{}$ and~$\MAGQ{}\sim\Dis{}$, i.e., we find that the torsion and non-metricity are just different parts of the distortion.
Any Lagrangian built from~$\MAGF{_{\mu\nu}^\rho_\sigma}$,~$\MAGT{_\mu^\alpha_\nu}$ and~$\MAGQ{_{\lambda\mu\nu}}$ can be decomposed in this way.
In other words, MAG can always be described as the theory of a rank-three \emph{matter} field~$\Dis{_\mu^\rho_\nu}$, which is coupled (minimally or otherwise) to a purely Riemannian theory of gravity, such as GR.
The mapping works both ways: any theory of a gravitating rank-three field is also a theory of MAG.
This realisation is important, because we have known for a long time that rank-three fields do not offer the best foundation upon which to build a quantum field theory (QFT), even in flat space.
It is especially hard to imagine how their various problems might be resolved by coupling them to gravity: this, in essence, is what the modern MAG project is trying to do.

\paragraph*{Too many particles} In a sufficiently flat spacetime, the field content of MAG comprises a symmetric rank-two metric perturbation~$\MAGH{_{\mu\nu}}\equiv\MAGg{_{\mu\nu}}-\tensor{\eta}{_{\mu\nu}}$, whilst the connection~$\MAGA{_\mu^\nu_\rho}$ is taken to be inherently perturbative. The connection poses an immediate problem, because it carries so many particles. Recall the definition of a massive particle from relativistic QFT, as a state of definite spin~$J$ and parity~$P$.\footnote{We consider parity-preserving theories, for which~$P$ will be a quantum number. In general, however, physical states can perfectly well be~$P$-indefinite.} Denoting these states by~$J^P_n$, where~$n$ allows for multiple degenerate representations, the massive little-group decomposition of the fields yields~\cite{Percacci:2020ddy,Marzo:2021esg}
\begin{subequations}
\begin{align}
\MAGA{_\mu^\nu_\rho} & \sim 3^-_1 \oplus 2^+_1 \oplus 2^+_2 \oplus 2^+_3 \oplus 2^-_1 \oplus 2^-_2 \oplus 1^+_1 \oplus 1^+_2
\nonumber\\ & \hspace{25pt}
\oplus 1^+_3 \oplus 1^-_1 \oplus 1^-_2 \oplus 1^-_3 \oplus 1^-_4 \oplus 1^-_5 \oplus 1^-_6
\nonumber\\ & \hspace{25pt}
\oplus 0^+_1 \oplus 0^+_2 \oplus 0^+_3 \oplus 0^+_4 \oplus 0^-_1  \, ,\label{DecompositionMAGA}
\\
\MAGH{_{\mu\nu}} & \sim 2^+_4 \oplus  1^-_7 \oplus 0^+_5 \oplus 0^+_6 \, .\label{DecompositionMAGH}
\end{align}
\end{subequations}
In the case of massless helicity states, the decomposition in~\cref{DecompositionMAGA,DecompositionMAGH} becomes even more crowded. This abundance of degrees of freedom (d.o.f), whose healthy propagation imposes contradictory requirements on the couplings, frustrates the building of stable and predictive quantum MAG models~\cite{Stelle:1977ry,Neville:1978bk,Neville:1979rb,Sezgin:1979zf,Hayashi:1979wj,Hayashi:1980qp,Barker:2023bmr}.

\begin{figure}[htbp]
\includegraphics[width=\linewidth]{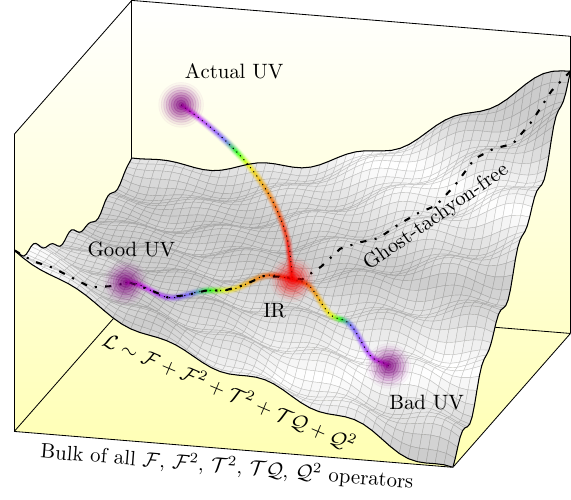}
\caption{\label{NoGoMAG} Ghost-tachyon-free metric-affine gravity (MAG) Lagrangians are normally made from tuned admixtures of the scalar invariants of the geometric tensors illustrated in~\cref{NonRiemannianSchematic}. Such theories will `de-tune' at the loop level, unless protected by symmetries beyond~$\DiffeomorphismGroup{}$ diffeomorphism invariance.}
\end{figure}

\paragraph*{Too few symmetries} We know from other sectors of physics that the way to consistently remove dangerous states is to impose gauge invariance.
In most of the modern literature, the basic symmetry of MAG is taken to be the~$\DiffeomorphismGroup{}$ diffeomorphism invariance\footnote{In fact, MAG was properly motivated as a gauge theory of the much larger rigid affine group~$\TranslationGroup{}\rtimes\GeneralLinearGroup{}$, the composition of the four-dimensional translations~$\TranslationGroup{}$ with the general linear transformations~$\GeneralLinearGroup{}$~\cite{Hehl:1994ue,NeemanRegge1978,ObukhovPereira2003}.
The metric~$\MAGg{_{\mu\nu}}$ and connection~$\MAGA{_\mu^\nu_\rho}$ are, however, gauge-invariant variables with respect to~$\GeneralLinearGroup{}$, such that only the~$\TranslationGroup{}$ symmetry remains visible, being manifest geometrically as~$\DiffeomorphismGroup{}$ invariance.
Something similar happens in the gauge theory of the Poincar\'{e} group~$\TranslationGroup\rtimes\LorentzGroup{}$, for which the tetrad and the spin connection gauge~$\TranslationGroup{}$ and~$\LorentzGroup{}$ respectively~\cite{Utiyama:1956sy,Kibble:1961ba,Sciama:1962}.
It is possible to reformulate Poincar\'{e} gauge theory (PGT) in terms of~$\MAGg{_{\mu\nu}}$ and~$\MAGA{_\mu^\nu_\rho}$ subject to the condition~$\MAGQ{_{\lambda\mu\nu}}\equiv 0$ in~\cref{MAGQDef} or, equivalently, in terms of~$\MAGT{_\mu^\alpha_\nu}$.
In these variables, it is local~$\LorentzGroup{}$ invariance which becomes hidden.
In the absence of fermionic matter, whose gravitational coupling requires tetrads,~$\DiffeomorphismGroup{}$-PGT makes precisely the same physical predictions as its more symmetric progenitor theory.
The same is true of MAG, which requires special kinds of non-Dirac spinors~\cite{SijackiNeeman1990,SijackiNeeman1992,NeemanSijacki1995}.
In both PGT and MAG, if the~$\DiffeomorphismGroup{}$ theory is sick, then the progenitor~$\TranslationGroup{}\rtimes\LorentzGroup{}$ or~$\TranslationGroup{}\rtimes\GeneralLinearGroup{}$ models cannot be healthy: this is one basic motivation for working with the~$\DiffeomorphismGroup{}$ formulations.
Another motivation is that the~$\DiffeomorphismGroup{}$ formulations offer a more familiar GR-type framework for developing the QFT, though there is a risk of overlooking quantum anomalies.
One can take a step further back and view~$\TranslationGroup{}\rtimes\LorentzGroup{}$-PGT as a local gauge theory of the full conformal group~$\ConformalGroup{}$ in scale-invariant variables: as shown in~\cite{Barker:2024goa}, this offers insights into the Yang--Mills structure that may help to guide the QFT of~$\DiffeomorphismGroup{}$-PGT.
Even if~$\TranslationGroup{}\rtimes\GeneralLinearGroup{}$-MAG offers similar insights, it appears to have been almost completely ignored in the modern MAG literature. Thus,~$\DiffeomorphismGroup{}$-MAG is the main focus of this paper.}
\begin{subequations}
\begin{align}
\tensor{\delta(h)}{_{\mu\nu}}&=\PD{_{\mu}}\Gen{_{\nu}}+\PD{_{\nu}}\Gen{_{\mu}}+\mathcal{O}(h),\label{DiffMAGH}\\
\tensor{\delta(A)}{_\mu^\rho_\nu}&=\PD{_{\mu}}\PD{_{\nu}}\Gen{^\rho}+\mathcal{O}(A).\label{DiffMAGA}
\end{align}
\end{subequations}
When consistently respected by all the non-linear interaction terms,~\cref{DiffMAGH,DiffMAGA} keep at bay the lower-spin ghosts\footnote{Interestingly, a smaller symmetry would also do, see~\cite{Alvarez:2005iy,Alvarez:2006uu,vanderBij:1981ym}.} while preserving the unique healthy propagation of the helicity-two graviton sector~$2^+_4$ in~$\MAGH{_{\mu\nu}}$~\cite{Deser:1969wk,Boulware:1974sr,Wald:1986bj}.
The~$\DiffeomorphismGroup{}$ gauge symmetry is, however, completely inadequate to remove the unwanted states from~$\MAGA{_\mu^\rho_\nu}$~\cite{Marzo:2021iok,Marzo:2024pyn}.
Countless studies have been dedicated to the selection of ghost-tachyon-free MAG Lagrangians at the linear level~\cite{Sezgin:1981xs,Blagojevic:1983zz,Blagojevic:1986dm,Kuhfuss:1986rb,Yo:1999ex,Yo:2001sy,Blagojevic:2002,Puetzfeld:2004yg,Yo:2006qs,Shie:2008ms,Nair:2008yh,Nikiforova:2009qr,Chen:2009at,Ni:2009fg,Baekler:2010fr,Ho:2011qn,Ho:2011xf,Ong:2013qja,Puetzfeld:2014sja,Karananas:2014pxa,Ni:2015poa,Ho:2015ulu,Karananas:2016ltn,Obukhov:2017pxa,Blagojevic:2017ssv,Blagojevic:2018dpz,Tseng:2018feo,Lin:2018awc,BeltranJimenez:2019acz,Zhang:2019mhd,Aoki:2019rvi,Zhang:2019xek,Jimenez:2019qjc,Lin:2019ugq,Percacci:2020ddy,Barker:2020gcp,BeltranJimenez:2020sqf,MaldonadoTorralba:2020mbh,Barker:2021oez,Marzo:2021esg,Marzo:2021iok,delaCruzDombriz:2021nrg,Baldazzi:2021kaf,Annala:2022gtl,Mikura:2023ruz,Mikura:2024mji,Barker:2024ydb}.
With very few exceptions~\cite{Barker:2024dhb}, this has been achieved by an arbitrary tuning of the couplings, not by the imposition of additional symmetries.
As illustrated in~\cref{NoGoMAG}, we will argue that this approach leads to fatal consequences stemming from the interacting QFT.
Interactions have always caused problems in MAG and related non-Riemannian theories, but these were so-far mostly studied in the context of classical field theory~\cite{Moller:1961,Pellegrini:1963,Hayashi:1967se,Cho:1975dh, Hayashi:1979qx,Hayashi:1979qx,Dimakis:1989az,Dimakis:1989ba,Lemke:1990su,Hecht:1990wn,Hecht:1991jh,Yo:2001sy,Afshordi:2006ad,Magueijo:2008sx,Charmousis:2008ce,Charmousis:2009tc,Papazoglou:2009fj,Baumann:2011dt,Baumann:2011dt,DAmico:2011eto,Gumrukcuoglu:2012aa,Wang:2017brl,Mazuet:2017rgq,BeltranJimenez:2020lee,JimenezCano:2021rlu,Barker:2022kdk,Delhom:2022vae,Annala:2022gtl,Barker:2022kdk,Barker:2023fem}, in the form of the strong coupling problem (see e.g.,~\cite{Vainshtein:1972sx,Deffayet:2001uk,Deffayet:2005ys,Charmousis:2008ce,Charmousis:2009tc,Papazoglou:2009fj,deRham:2014zqa,Deser:2014hga,Wang:2017brl,Barker:2022kdk}) and broken `accidental' symmetries~\cite{Velo:1969txo,Aragone:1971kh,Cheng:1988zg,Hecht:1996np,Chen:1998ad,Yo:1999ex,Yo:2001sy,Blixt:2018znp,Blixt:2019ene,Blixt:2020ekl,Krasnov:2021zen,Bahamonde:2021gfp,Delhom:2022vae,Karananas:2024hoh,Karananas:2024qrz} (see also~\cite{Hayashi:1980qp,Blagojevic:1983zz,Blagojevic:1986dm,Yo:2001sy,Blagojevic:2002,Ong:2013qja,Blagojevic:2013dea,Blagojevic:2013taa,Blagojevic:2018dpz,BeltranJimenez:2019hrm,Aoki:2020rae,Barker:2021oez}). 
What, then, are the quantum implications of interactions?
A general answer to this question, about the systematics of renormalisation, is already well developed in the form of effective field theory (EFT).
In turn, EFT methods tell us that \emph{symmetries} should be the guiding principle for model-building.
The popular~$\DiffeomorphismGroup{}$ version of MAG introduces many new d.o.f, but it runs into trouble because it does not `pay for them' with additional symmetries.
We already demonstrated that~$\DiffeomorphismGroup{}$-MAG is really nothing more than a theory of higher-rank matter coupled to a Riemannian gravity theory.
For this Riemannian sector, GR itself provides the `gold standard' foundation upon which to build a gravitational EFT~\cite{Donoghue:1994dn}. Most $\DiffeomorphismGroup{}$-MAG models in the literature contain a copy of GR embedded within the metric-affine Einstein--Hilbert term~$\mathcal{L}\sim\MAGF{}$, though this operator is not motivated by EFT principles, as we will now see.

\paragraph*{Einstein--Hilbert appropriation} Models of the form~$\mathcal{L}\sim\MAGF{}$ or~$\mathcal{L}\sim\MAGF{}+\MAGT{}^2+\MAGQ{}^2$ are similar to GR: in the absence of matter, the distortion integrates out as a non-dynamical field, leaving behind the unique model of pure GR, which is defined as~$\mathcal{L}\sim\MAGR{}$. With matter, this same process leads to effective interactions whose effects are explored in (e.g.)~\cite{Rigouzzo:2022yan,Rigouzzo:2023sbb}.
This model-building approach sounds compelling, but there are important differences in predictivity and physical motivation between GR, for which~$\mathcal{L}\sim\MAGR{}$, and the Einstein--Hilbert action~$\mathcal{L}\sim\MAGF{}$ in $\DiffeomorphismGroup{}$-MAG. As mentioned above, the latter is sometimes put forward as a theory of gravity (when assuming~$\MAGQ{}\equiv 0$ this corresponds to Einstein--Cartan theory). Unlike for GR, however,~$\mathcal{L}\sim\MAGF{}$ is posited purely on the geometric grounds of~\cref{NonRiemannianSchematic} and cannot be justified by any EFT principle; even the so-called \emph{projective} symmetry of~$\mathcal{L}\sim\MAGF{}$ is not strong enough to make it unique at the lowest derivative order~\cite{Barker:2024dhb}. This contrasts strongly with GR: whilst geometric principles akin to~\cref{NonRiemannianSchematic} undoubtedly served as Einstein's original inspiration~\cite{einstein1916grundlage,ricci1900,christoffel1869,riemann1868,gauss1828}, the essential (i.e. physically significant) aspect that was understood much later is that~$\mathcal{L}\sim\MAGR{}$ is the \emph{unique} low-energy limit of~$\DiffeomorphismGroup{}$ gauge theory as an EFT~\cite{Boulware:1974sr,Deser:1969wk,Deser:1987uk,Wald:1986bj}.\footnote{The quantum ramifications are best known from the works of Donoghue, see~\cite{Donoghue:1994dn}. The $\DiffeomorphismGroup{}$ bootstrap was developed by Deser; see discussions of potential ambiguity in this bootstrapping process in~\cite{Padmanabhan:2004xk}, apparently resolved in~\cite{Butcher:2009ta}, and attempts to extend it to MAG in~\cite{Delhom:2022zbi}.} Because they lack an EFT basis, it is unclear how Einstein--Hilbert-type $\DiffeomorphismGroup{}$-MAG theories can be truly predictive, though they have the advantage of being extremely simple in terms of classical manipulations.\footnote{Note however that one serious branch of the literature does put Einstein--Hilbert-type models forwards as consistent quantum theories, see~\cite{Bauer:2010jg,Shaposhnikov:2020aen,Shaposhnikov:2020geh,Shaposhnikov:2020fdv,Karananas:2021zkl,Rigouzzo:2022yan,Rigouzzo:2023sbb}, and recent progress in~\cite{Karananas:2025ews,Shaposhnikov:2025znm}. The status of such models as infrared foundations will be investigated elsewhere.}

\paragraph*{Validity of effective field theory} Working still in the geometric basis of operators from~\cref{NonRiemannianSchematic}, the real problems of inconsistent spectra only arise when extending the Einstein--Hilbert term to models of the form~$\mathcal{L}\sim\MAGF{}+\MAGF{}^2+\MAGT{}^2+\MAGQ{}^2$. These problems should be resolved by EFT methods. From the outset, it is natural to ask whether EFT principles, familiar from the standard model effective field theory (SMEFT) of particle physics, can be safely applied to gravity without special care or modification to the physics. The answer is affirmative. Indeed, when applied to `gravity proper' --- i.e. to GR --- the EFT framework known as GREFT is well established and highly regarded~\cite{Donoghue:2019jeq,Donoghue:2017ovt,Donoghue:2015hwa,Donoghue:2012zc}. A central tenet of GREFT is that the masslessness of the graviton is protected by gauge invariance at all orders of perturbation theory. Thus, the lowest energy at which GREFT becomes applicable is already zero, and the theory is expected to remain perturbative all the way up to the Planck scale at $\MPl{}\sim\SI{e19}{\giga\electronvolt}$. This is as well, because the characteristic energy at which we study gravitational physics is very tiny: gravitational wave astronomy has so far been commonly conducted at up to $\sim\SI{e2}{\hertz}$, which by loose analogy to collider physics translates to $\sqrt{s}\sim\SI{e-12}{\electronvolt}$.\footnote{Of course $\sqrt{s}$ has no meaningful interpretation here in terms of a scattering of quanta, but the separation of scales is numerically valid.} So much for GR: what about \emph{new} d.o.f that might accompany a modified gravity model such as MAG? Observational constraints dictate that any such species cannot be massless in the same way as the graviton, so their inclusion in a phenomenologically useful EFT of MAG is expected to only become distinguishable from GREFT above some threshold energy, whose value depends on the ultraviolet (UV) completion and is connected to the propagation of the new d.o.f. The crucial point, however, is that the symmetric model-building paradigm of EFT is already mandated long before any such threshold energy has been specified. This is reflected in the developmental history of the standard model of particle physics, which ab initio lacks any reference to such a threshold scale in its symmetry-derived base formulation --- its \emph{infrared (IR) foundation}. The symmetric IR foundation is a necessary first step in developing the full standard model, in which scales are acquired through various quantum mechanisms such as quark confinement, and chiral/electroweak symmetry breaking. Despite the name of the latter, all such processes are fundamentally covariant. Indeed, it is a champion result of theoretical physics that the full standard model, complete with scales, remains renormalizable precisely because the symmetries of the IR foundation and its BRST identities remain intact. Such effects do alter the true low-energy d.o.f and these call for a second round of `phenomenological' EFT model-building in order to be studied. This two-step process has been hugely successful, and we suggest that it tells us what we can reasonably expect from MAG. Using EFT principles, we will therefore not be surprised to find that all the IR foundations of MAG (we show that there are several) feature new low-energy d.o.f that are \emph{exclusively massless}, just like the graviton itself. And yet, this tells us \emph{very little} about the new polarisations that we might expect to see in gravitational wave astronomy, and \emph{absolutely nothing} about the threshold energy at which we might expect to see them. The theoretical development of MAG in general, and the problem of the mass acquisition of MAG in particular, are nowhere near advanced enough for such predictions to be made. Indeed, serious EFT-based endeavours in the theoretical development of MAG have only recently been revived by a sub-set of authors~\cite{Baldazzi:2021kaf,Martini:2023apm,Sauro:2022chz,Sauro:2024ujx,Melichev:2023lwj,Melichev:2025hcg}, and it is in this spirit that we try to proceed.

\paragraph*{In this work} We restrict our analysis in two ways. Firstly, we consider the special case of MAG in which the distortion defined in~\cref{postriem} is fully symmetric\footnote{See e.g.~\cite{Iosifidis:2023wbx} for an example of where this restriction is also made, with analogies being drawn to statistical manifolds.}
\begin{equation}\label{SymmetricDistortion}
\Dis{_{\mu\nu\rho}}\equiv\Dis{_{(\mu\nu\rho)}} \,.
\end{equation}
From~\cref{MAGTDef}, the restriction in~\cref{SymmetricDistortion} implies~$\MAGT{_\mu^\alpha_\nu}\equiv 0$ by construction.
The spacetime thus has curvature and (somewhat restricted) non-metricity, but no torsion.
Secondly, we consider only parity-preserving models.
This limitation has no physical grounding (we know already that parity is maximally violated in the weak sector, and there are no phenomenological reasons to expect parity-preservation in gravity), so like~\cref{SymmetricDistortion} it just serves to simplify the initial analysis.
Our objective is to identify the possible quantum foundations for MAG, by searching for models whose freedom from ghosts and tachyons is fully guaranteed by symmetries, rather than by the tuning of couplings. For a preliminary analysis, we make the restriction to so-called \emph{free} symmetries, whose generators are (possibly symmetrized) tensor fields without extra constraints such as having vanishing divergence (this condition will be motivated in~\cref{Sec:Stab}). We find two such quantum configurations, each propagating a single massless mode of different spin:
\begin{description}
\item[Spin-one] In order to propagate a QED-type massless vector, it seems natural to use a scalar symmetry generator. We find, however, that scalar generators lead only to sick models, and a tensor generator is instead required. The non-linear completion of the theory corresponds to the well-known operator which is the square of the homothetic curvature.
\item[Spin-three] Our algorithm leads us by five different paths to Fronsdal theory. Four of these paths arise from considering tensor generators, consistent with the usual traceless rank-two Fronsdal generator. The fifth route arises, somewhat unusually, from considering a vector generator.
\end{description}
The paper is set out as follows. In~\cref{Sec:Pred} we motivate the symmetry-first approach with a pedagogical discussion of EFT, making the connection to gravity clear. The algorithm used is presented in~\cref{Sec:Stab}, in particular we motivate our ansatz for the subsequent analysis, which we give in~\cref{VeryFlatLag}. The results are presented in~\cref{Results}. In~\cref{Sec:Concl} we summarise our conclusions, in particular reiterating how (as argued above) the masslessness of the new species is not an important phenomenological quality at this nascent stage in the development of MAG as a theory of physics. Thus, the major science products are the classification of the IR foundations, their na\"ive spectra and symmetries.

\paragraph*{Conventions} We adopt natural units in which~$\hbar\equiv c\equiv 1$, and the particle physics signature~$(+,-,-,-)$. As far as possible we adhere to the MAG conventions set out in~\cite{Percacci:2020ddy}.

\section{Effective field theory}\label{Sec:Pred}

\paragraph*{General idea} In this section we examine the assumptions that must be made about MAG as an EFT.
EFT methods have shaped our understanding of how physical scales and quantum fluctuations organize in manifesting scale-decoupling, and the causal and local phenomenology~\cite{Georgi:1993mps,Burgess:2007pt,Bijnens:2006zp,Manohar:1996cq,Kaplan:1995uv,Pich:1995bw,Ecker:1994gg,Dobado:1989ax,Dobado:1989gr,Weinberg:1978kz,Coleman:1969sm,Callan:1969sn}.
The term `\emph{effective}' implies the ability to draw unambiguous and universal predictions.
EFTs extend our domain of predictive field models to include, on top of the power-counting renormalizable sector~$\mathcal{L}_{n\leq 4}$, an infinite tower of operators with increasing mass dimension
\begin{align}\label{EFTLag}
\mathcal{L}=\mathcal{L}_{n\leq4}+\sum_{n=5}^{\infty}\sum_i\frac{\GeneralG{n}{_i}\OperatorO{i}{n}}{\Lambda^{n-4}} \,,
\end{align}
where the~$\big\{\GeneralG{n}{_i}\big\}$ are dimensionless couplings\footnote{We will replace~$\alpha$ with a new symbol for each specific theory.} with~$i$ labelling multiple operators of given mass dimension~$n$ (if any). The constant~$\Lambda$ is a cutoff scale, assumed to be larger than the experimental energy~$E$ we wish to probe. 
The inclusive appearance of~\cref{EFTLag} can deceive one into thinking that EFTs are simply defined by \emph{not being} power-counting renormalizable: this is not the case.
Indeed, as shown for instance in~\cite{Gomis:1995jp}, the operators in all parts of~\cref{EFTLag} are strongly constrained so as to support a sensible perturbative expansion.

\paragraph*{Strictly renormalisable theories} In a strictly renormalisable theory we are limited in~\cref{EFTLag} to~$\mathcal{L}_{n \leq 4}$, along with propagators behaving as~$1/k^2$ in the large momentum limit.
A small-coupling perturbative expansion is performed in parallel with an expansion in an increasing number of loops.
Confronted with an experimental determination of given precision~$\Delta_{\text{Ex}}$, the theorist must provide predictions with a theoretical error~$\Delta_{\text{Th}} < \Delta_{\text{Ex}}$.
Loops enter as soon as~$\Delta_{\text{Ex}}$ is sufficiently small, resulting in an unphysical dependence on the unknown UV behaviour.
A small collection of low energy constants (LECs) is then required to conceal this dependence through a procedure of matching against experimental measurements~\cite{Ecker:1994gg}.

\paragraph*{Yukawa theory} We illustrate this with Yukawa theory, taking~$\big\{\YukawaG{n}{_i}\big\}$ as notation for the set of couplings in~\cref{EFTLag}.
We consider a scalar boson~$\Phi$ of square mass~$\YukawaG{2}{}\Lambda^2$, and a Dirac fermion~$\psi$ of mass~$\YukawaG{3}{}\Lambda$, coupled by a simple Yukawa-type interaction with dimensionless coupling~$\YukawaG{4}{}$, such that\footnote{In principle, the kinetic interactions in~\cref{Yukawa} could also be parameterised by couplings, but we absorb these canonically into field normalisations.}
\begin{equation}\label{Yukawa}
\mathcal{L}=\PD{_\mu}\Phi\PD{^{\mu}}\Phi - \YukawaG{2}{}\Lambda^2\Phi^2 + \bar\psi \slashed{\partial}\psi - \YukawaG{3}{}\Lambda\bar \psi\psi + \YukawaG{4}{} \Phi \bar \psi\psi \,.
\end{equation}
We consider the paradigmatic decay of the scalar into a lighter fermion-antifermion pair, as measured by a precise-enough experiment that its modelling with~\cref{Yukawa} requires loop correction. A schematic representation of the components of the corrected amplitude components is:
\begin{align*}
\resizebox{\linewidth}{!}{
\begin{tikzpicture}
\begin{feynman}
\draw[fill=black] (5,0) circle (0.12) node[above=5pt]; 
\vertex (plus1) at (0, -3.5)  {\( \scalebox{2}{$\YukawaG{4}{}$} \)};
\vertex (plus2) at (5, -3.5)  {\( \scalebox{2}{$\delta\YukawaG{4}{}$} \)};
\vertex (plus3) at (10, -3.5)  {\( \scalebox{2}{$ \frac{1}{\epsilon} + \log\left(\frac{\Lambda^2}{\mu^2}\right)+\dots$  } \)};
\vertex (b0l) at (-2.3, -2.3);
\vertex (c0l) at (2.3, -2.3);
\vertex (d0l) at (0.5, 3.);
\vertex (a0) at (0, 0);
\vertex (b0) at (2, 2) {\( \scalebox{2}{$\psi$} \)};
\vertex (c0) at (2, -2) {\( \scalebox{2}{$\bar\psi$} \)};
\vertex (d0) at (-2, 0) {\( \scalebox{2}{$\Phi$} \)};
\vertex (a1) at (5, 0);
\vertex (b1) at (7, 2) {\( \scalebox{2}{$\psi$} \)};
\vertex (c1) at (7, -2) {\( \scalebox{2}{$\bar\psi$} \)};
\vertex (d1) at (3, 0) {\( \scalebox{2}{$\Phi$} \)};
\vertex (a2) at (9, 0);
\vertex (b2) at (12, 2) {\( \scalebox{2}{$\psi$} \)};
\vertex (c2) at (12, -2) {\( \scalebox{2}{$\bar\psi$} \)};
\vertex (d2) at (8, 0) {\( \scalebox{2}{$\Phi$} \)};
\vertex (xL2) at (11, 1);
\vertex (xR2) at (11, -1);
\diagram* {
(d0) -- [scalar, black, ultra thick] (a0),
(b0) -- [fermion, black ] (a0),
(a0) -- [fermion, black] (c0),
(d1) -- [scalar, black, ultra thick] (a1),
(b1) -- [fermion, black ] (a1),
(a1) -- [fermion, black] (c1) ,
(d2) -- [scalar, black, ultra thick] (a2),
(b2) -- [fermion, black ] (xL2),
(xL2) -- [fermion, black ] (a2),
(a2) -- [fermion, black] (xR2),
(xR2) -- [fermion, black] (c2),
(xL2) -- [scalar, black, ultra thick] (xR2),
};
\end{feynman}
\end{tikzpicture}
}
\end{align*}
To render calculations finite, the spacetime is shifted away from four dimensions by a small regulator~$\epsilon$, and powers of the fiducial renormalisation scale~$\mu$ are simultaneously introduced in order to maintain dimensional consistency.
The coupling~$\YukawaG{4}{}$ runs so as to cancel the unphysical dependence of the prediction on~$\mu$, and it is also shifted by a counterterm~$\delta\YukawaG{4}{}$, which absorbs the~$1/\epsilon$ divergence as the regulator is removed.
In this simple example,~$\YukawaG{4}{}$ is the LEC to be fixed with a direct comparison with the experiment.
It is a highly non-trivial result that models exist requiring only a finite number of LECs and so a finite number of experimental inputs.
Once the LECs are determined, every other independent process belongs to the predictions of the theory.
Faced with increasingly smaller~$\Delta_{\text{Ex}}$, one needs only to recompute the LECs to higher loop order.

\paragraph*{Generally renormalisable theories} The inclusion of operators of mass dimension greater than four, the default scenario for gravitational theories, appears to threaten predictivity.
These operators lead to more strongly divergent loop integrals.
The counterterms required to absorb these divergences can only come from operators of yet higher mass dimension.
Consequently, starting with a restricted set~$\big\{\OperatorO{i}{n}\big\}$, one is compelled to include an infinite tower of them, parameterised by infinitely many couplings as in~\cref{EFTLag}.
This procedure `closes' in a consistent manner only when all the operators are subject to symmetry constraints, resulting in a generally renormalisable theory~\cite{Gomis:1995jp,Gasser:1983yg,Bijnens:2006zp,Pich:1995bw,Ecker:1994gg,Gasser:1984gg,Bijnens:1997vq,Colangelo:2001df,Bijnens:1998yu}.
We can truncate this infinite tower, so long as we maintain control over the neglected part by using a double perturbative expansion in the~$\big\{\GeneralG{n}{_i}\big\}$ and~$E/\Lambda$.
The theory truncated at order~$n$ can be used to model experiments for which~$(E/\Lambda)^{n-4} \sim \Delta_{\text{Ex}}$.
Predictivity is thus restored until~$E$ approaches~$\Lambda$, where perturbativity breaks down and new d.o.f are expected to shape the dynamics.

\paragraph*{Chiral perturbation theory} To clarify the differences between strictly and generally renormalizable theories, we consider pions.
These have been used as gravity analogues previously~\cite{Weinberg:1978kz}, but the results were not previously presented in the modern dimensional regularisation scheme, which we use here.\footnote{See~\cite{Marzo:2024pyn} for a similar example case of Abelian vector EFTs.}
A triplet of pion fields~$\vec{\pi}\equiv(\pi_1, \pi_2, \pi_3)$ can be used to describe the propagating massless Goldstone bosons which survive spontaneous breaking of the (chiral, in its action over fermion states) symmetry~$\mathrm{SU}(2)_\text{L}\times \mathrm{SU}(2)_\text{R}$. Using a notation~$\big\{\ChiralG{n}{_i}\big\}$ for the set of couplings, the chiral Lagrangian is written  
\begin{align}\label{ChPt}
\mathcal{L} & = \ChiralG{4}{}\ChiralD{_{\mu}} \vec{\pi} \cdot \ChiralD{^{\mu}} \vec{\pi} 
+ \frac{\ChiralG{8}{_1}}{\Lambda^4} (\ChiralD{_{\mu}} \vec{\pi} \cdot \ChiralD{^{\mu}} \vec{\pi})
(\ChiralD{_{\nu}} \vec{\pi} \cdot \ChiralD{^{\nu}} \vec{\pi})
\nonumber\\& \hspace{15pt}
+ \frac{\ChiralG{8}{_2}}{\Lambda^4}
(\ChiralD{_{\mu}} \vec{\pi} \cdot \ChiralD{_{\nu}} \vec{\pi})
(\ChiralD{^{\mu}} \vec{\pi} \cdot \ChiralD{^{\nu}} \vec{\pi})
+ \dots,
\end{align}
where the derivative is defined as\footnote{The structure of this chiral-covariant derivative is motivated in~\cite{Weinberg:1968de}. The components~$\tensor{\Psi}{_i}$ in some representation of a chiral-covariant field~$\Psi$ transform under the axial generators~$\tensor{X}{_a}$ of~$\mathrm{SU}(2)_\text{L}\times \mathrm{SU}(2)_\text{R}$ as~$\left[\tensor{X}{_a}, \tensor{\Psi}{_i}\right] = - \tensor{\epsilon}{_{a b c}} \left(\tensor{\pi}{^b}/\Lambda\right)\tensor{t}{^c_{ij}}\tensor{\Psi}{_j}$, with~$\tensor{t}{^c_{ij}}$ the components of a Hermitian matrix defining the representation of~$\Psi$ and~$\tensor{\pi}{^a}$ the components of~$\vec{\pi}$. Invariants can then be built via standard index contraction and conjugate transposition of these~$\tensor{\Psi}{_i}$ components. The pions themselves are not chiral-covariant: their non-linear transformations follow~$\left[\tensor{X}{_a}, \tensor{\pi}{_b}\right] = - i \Lambda \left(\tfrac{1}{2}\left(1 - \vec{\pi}\cdot\vec{\pi}/\Lambda^2\right)\tensor{\delta}{_{ab}} + \tensor{\pi}{_a} \tensor{\pi}{_b} /\Lambda^2\right)$. The point of~\cref{ChiralD} is to non-linearly combine the pions so as to form a chiral-covariant object~$\ChiralD{_{\mu}} \vec{\pi}$ in the~$\tensor{t}{^a_{b c}} = - i \tensor{\epsilon}{^a_{b c}}$ representation --- albeit one with an extra spacetime index that must be contracted away when constructing Lorentz-scalar operators.}
\begin{align}\label{ChiralD}
\ChiralD{_{\mu}} \vec{\pi} \equiv \frac{\PD{_{\mu}} \vec{\pi}}{1 + \vec{\pi}\cdot\vec{\pi}/\Lambda^2} \,,
\end{align}
and the cutoff~$\Lambda\sim\SI{e2}{\mega\electronvolt}$ is set by pion decay.
The series in~\cref{ChPt} continues at~$n$th order in~$\ChiralD{_\mu}$ with coefficients~$\big\{\ChiralG{n}{_i}/\Lambda^{n-4}\big\}$, subject to the only constraint that chiral symmetry is imposed at each order.
Consider the pion-pion scattering amplitude at tree level:
\begin{equation*}
\resizebox{\linewidth}{!}{%
\begin{tikzpicture}
\begin{feynman}
\vertex (bound1) at (-2, 0) {\( \scalebox{2}{$$} \)};
\vertex (bound2) at (12, 0) {\( \scalebox{2}{$$} \)};
\vertex (a0) at (5, 0);
\vertex (b0) at (7, 2) {\( \scalebox{2}{$\pi_c$} \)};
\vertex (c0) at (7, -2) {\( \scalebox{2}{$\pi_d$} \)};
\vertex (d0) at (3, 2) {\( \scalebox{2}{$\pi_b$} \)};
\vertex (e0) at (3, -2) {\( \scalebox{2}{$\pi_a$} \)};
\diagram* {
(a0) -- [scalar, black, ultra thick] (b0),
(a0) -- [scalar, black, ultra thick] (c0),
(d0) -- [scalar, black, ultra thick] (a0),
(e0) -- [scalar, black, ultra thick] (a0),
};
\end{feynman}
\end{tikzpicture}%
}
\end{equation*}
From~\cref{ChPt}, this has the contributions in terms of Mandelshtam variables~$s$,~$t$ and~$u$
\begin{align}\label{amp0}
\mathcal{M}^{0\text{L}}& =\big(1/\ChiralG{4}{}^{2}\Lambda^{2}\big)\Big[\Kronecker{_{ab}}\Kronecker{_{cd}}\,\big[4\ChiralG{4}{}(s-t-u)
\nonumber\\ &\hspace{-25pt}
-\big(2\ChiralG{8}{_1}/\Lambda^{2}\big)s(t+u)
-\big(\ChiralG{8}{_2}/\Lambda^{2}\big)\big(s(t+u)+2tu\big)\big]
\nonumber\\ &\hspace{-25pt}
-\Kronecker{_{ad}}\Kronecker{_{bc}}\,\big[4\ChiralG{4}{}(s+t-u)
+\big(2\ChiralG{8}{_1}/\Lambda^{2}\big)u(s+t)
\nonumber\\ &\hspace{-25pt}
+\big(\ChiralG{8}{_2}/\Lambda^{2}\big)\big(2st+su+tu\big)\big]
-\Kronecker{_{ac}}\Kronecker{_{bd}}\,\big[4\ChiralG{4}{}(s-t+u)
\nonumber\\ &\hspace{-25pt}
+\big(2\ChiralG{8}{_1}/\Lambda^{2}\big)t(s+u)
+\big(\ChiralG{8}{_2}/\Lambda^{2}\big)\big(st+2su+tu\big)\big]\Big]\,.
\end{align}
It is evident from~\cref{amp0} that the leading-order effects depend only on~$\ChiralG{4}{}$, so the theory is predictive up to~$\mathcal{O}(E^2/\Lambda^2)$.
For smaller~$\Delta_{\text{Th}}$, one admits the one-loop corrections:
\begin{align*}
\resizebox{\linewidth}{!}{%
\begin{tikzpicture}
\begin{feynman}
\vertex (a0p) at (1, 0);
\vertex (a0m) at (-1, 0);
\vertex (b0) at (2, 2) {\( \scalebox{2}{$\pi_c$} \)};
\vertex (c0) at (2, -2) {\( \scalebox{2}{$\pi_d$} \)};
\vertex (d0) at (-2, 2) {\( \scalebox{2}{$\pi_b$} \)};
\vertex (e0) at (-2, -2) {\( \scalebox{2}{$\pi_a$} \)};
\vertex (a1p) at (5, 1);
\vertex (a1m) at (5, -1);
\vertex (b1) at (7, 2) {\( \scalebox{2}{$\pi_c$} \)};
\vertex (c1) at (7, -2) {\( \scalebox{2}{$\pi_d$} \)};
\vertex (d1) at (3, 2) {\( \scalebox{2}{$\pi_b$} \)};
\vertex (e1) at (3, -2) {\( \scalebox{2}{$\pi_a$} \)};
\vertex (a2p) at (10, 1);
\vertex (a2m) at (10, -1);
\vertex (b2) at (12, 2) {\( \scalebox{2}{$\pi_c$} \)};
\vertex (c2) at (12, -2) {\( \scalebox{2}{$\pi_d$} \)};
\vertex (d2) at (8, 2) {\( \scalebox{2}{$\pi_b$} \)};
\vertex (e2) at (8, -2) {\( \scalebox{2}{$\pi_a$} \)};
\diagram* {
 (a0p) -- [scalar, black, ultra thick] (b0),
 (a0p) -- [scalar, black, ultra thick] (c0),
 (a0m) -- [scalar, black, ultra thick] (d0),
 (a0m) -- [scalar, black, ultra thick] (e0),
 (a0p) -- [half left, scalar, black, ultra thick] (a0m),
 (a0p) -- [half right, scalar, black, ultra thick] (a0m),
 (a1p) -- [scalar, black, ultra thick] (b1),
 (a1m) -- [scalar, black, ultra thick] (c1),
 (a1p) -- [scalar, black, ultra thick] (d1),
 (a1m) -- [scalar, black, ultra thick] (e1),
 (a1p) -- [half left, scalar, black, ultra thick] (a1m),
 (a1p) -- [half right, scalar, black, ultra thick] (a1m),
 (a2p) -- [scalar, black, ultra thick] (c2),
 (a2m) -- [scalar, black, ultra thick] (b2),
 (a2p) -- [scalar, black, ultra thick] (d2),
 (a2m) -- [scalar, black, ultra thick] (e2),
 (a2p) -- [half left, scalar, black, ultra thick] (a2m),
 (a2p) -- [half right, scalar, black, ultra thick] (a2m),
};
\end{feynman}
\end{tikzpicture}%
}
\end{align*}
The dominant part of the vertex also dominates in the loop. Computing the relevant integrals, and expanding the two-point Passarino--Veltman scalar function (see~\cite{Passarino:1978jh,Shtabovenko:2023idz,Shtabovenko:2016sxi,Shtabovenko:2016whf,Mertig:1990an,Carrazza:2016gav,Patel:2015tea}) in terms of the regulator~$\epsilon$, leads to
\begin{align}\label{amp2}
\mathcal{M}^{1\text{L}}&=\left(1/3\pi^2\epsilon\ChiralG{4}{}^{2}\Lambda^{4}\right)\Big[\Kronecker{_{ab}}\Kronecker{_{cd}}\big[-5s(t+u)+3t^2
\nonumber\\&\hspace{10pt}
-2tu+3u^2\big]
+\Kronecker{_{ad}}\Kronecker{_{bc}}\big[3s^2-5u(s+t)
\nonumber\\&\hspace{10pt}
-2st+3t^2\big]
+\Kronecker{_{ac}}\Kronecker{_{bd}}\big[3s^2-5st-2su
\nonumber\\&\hspace{10pt}
-5tu+3u^2\big]\Big]
+\mathcal{O}\left(E^6/\Lambda^6\right)
\,.
\end{align}
It is clear from the quadratic dependence on the Mandelstam variables in~\cref{amp2} that, while the main contribution to the scattering is defined by~$\ChiralG{4}{}$, its corrections cannot be renormalised through a rescaling of the same parameter.
This role is instead played by the~$\big\{\ChiralG{8}{_i}\big\}$, and a close comparison of~\cref{amp0,amp2} shows that the infinity can be absorbed by the counterterms
\begin{align} \label{ampct}
\delta\ChiralG{8}{_1}\equiv-\frac{2}{3\pi^2\epsilon},\quad\delta\ChiralG{8}{_2}\equiv-\frac{4}{3\pi^2\epsilon}\,.
\end{align}
Thus, decreasing~$\Delta_{\text{Ex}}$ requires the continuous introduction of new LECs such as the~$\big\{\ChiralG{8}{_i}\big\}$, which must themselves be supported by further independent experimental inputs.
Only if \emph{all} chirally symmetric operators are included in~\cref{ChPt}, can we be sure that we have a counterterm for each divergence.
Interestingly, however, the new LECs~$\big\{\ChiralG{8}{_i}\big\}$ do not determine the non-analytic part of the scattering amplitude, which encodes a purely quantum effect through the branch-cut singularities
\begin{align}\label{lowt}
&\mathcal{M}^{1\text{L}}_{\text{UV}}=\left(2/3\pi^2\ChiralG{4}{}^2\Lambda^4\right)
\bigg[\Kronecker{_{ab}}\Kronecker{_{cd}}\Big[3s^2\log\left(-\mu^2/s\right)
\nonumber\\&\hspace{10pt}
+(s+2t)\big[
(s+t)\log\left(\mu^2/(s+t)\right)
+t\log\left(-\mu^2/t\right)
\big]\Big]
\nonumber\\&\hspace{10pt}
+\Kronecker{_{ad}}\Kronecker{_{bc}}\Big[
(s-t)\big[
s\log\left(-\mu^2/s\right)
-t\log\left(-\mu^2/t\right)
\big]
\nonumber\\&\hspace{10pt}
+3(s+t)^2\log\left(\mu^2/(s+t)\right)
\Big]
\nonumber\\&\hspace{10pt}
+\Kronecker{_{ac}}\Kronecker{_{bd}}\Big[
3t^2\log\left(-\mu^2/t\right)
+s(2s+t)\log\left(-\mu^2/s\right)
\nonumber\\&\hspace{10pt}
+(s+t)(2s+t)\log\left(\mu^2/(s+t)\right)
\Big]
\bigg] \,.
\end{align}
The singular behaviour of~\cref{lowt} describes the universal low-energy part of the scattering process. As stressed in~\cite{Weinberg:1978kz}, it is separated from the unknown high-energy completion and can be inferred by the general constraints of analyticity, causality and chiral symmetry. Low-energy theorems such as~\cref{lowt} are a major phenomenological product of EFTs.

\paragraph*{General relativity} Chiral perturbation theory is closely analogous to the EFT extension of GR~\cite{Donoghue:2019jeq,Donoghue:2017ovt,Donoghue:2015hwa,Donoghue:2012zc}. The latter fits perfectly into~\cref{EFTLag}, though it is easiest to see this by first rescaling the fields. In particular, the metric perturbation~$\MAGH{_{\mu\nu}}$ introduced in~\cref{DecompositionMAGH} is dimensionless, consistent with the usual treatments in linear gravity and cosmological perturbation theory; in EFT terms, however, a propagating bosonic field such as the graviton should really have an engineering dimension of one.
Thus, working in the standard perturbation scheme defined by
\begin{equation}\label{PerturbativeExpansion}
\MAGg{_{\mu\nu}}\equiv \G{_{\mu\nu}}+\MAGH{_{\mu\nu}},
\quad
\MAGg{^{\mu\nu}}= \G{^{\mu\nu}}-\MAGH{^{\mu\nu}}+\mathcal{O}\left(\MAGH{}^2\right),
\end{equation}
we introduce a canonically rescaled field
\begin{equation}\label{MAGEFTHDef}
\MAGEFTH{_{\mu\nu}} \equiv \Lambda\MAGH{_{\mu\nu}},
\end{equation}
where the cutoff scale~$\Lambda\sim\SI{e19}{\giga\electronvolt}$ is proportional to Planck mass.
Under the scheme in~\cref{PerturbativeExpansion} and in terms of the rescaled field in~\cref{MAGEFTHDef}, the Einstein--Hilbert operator reads
\begin{align}
\sqrt{-g}\MAGR{}&=
\frac{1}{\Lambda}\bigg[\PD{_\alpha}\PD{_\beta}\MAGEFTH{^{\alpha\beta}}
-\PD{_\alpha}\PD{^\alpha}\MAGEFTH{}\bigg]
+\frac{1}{\Lambda^2}\bigg[
\MAGEFTH{^{\alpha\beta}}\PD{_\alpha}\PD{_\beta}\MAGEFTH{}
\nonumber\\ &\hspace{-30pt}
-\MAGEFTH{^{\alpha\beta}}\PD{_\chi}\PD{_\beta}\MAGEFTH{^\chi_\alpha}
-\frac{1}{4}\PD{_\beta}\MAGEFTH{}\PD{^\beta}\MAGEFTH{}
-\PD{_\alpha}\MAGEFTH{^{\alpha\beta}}\PD{_\chi}\MAGEFTH{_\beta^\chi}
\nonumber\\ &\hspace{-30pt}
+\PD{^\beta}\MAGEFTH{}\PD{_\chi}\MAGEFTH{_\beta^\chi}
-\MAGEFTH{^{\alpha\beta}}\PD{_\beta}\PD{_\chi}\MAGEFTH{_\alpha^\chi}
+\frac{1}{2}\MAGEFTH{}\PD{^\beta}\PD{_\chi}\MAGEFTH{^\beta^\chi}
\nonumber\\ &\hspace{-30pt}
+\MAGEFTH{^{\alpha\beta}}\PD{_\chi}\PD{^\chi}\MAGEFTH{_{\alpha\beta}}
-\frac{1}{2}\MAGEFTH{}\PD{_\chi}\PD{^\chi}\MAGEFTH{}
-\frac{1}{2}\PD{_\beta}\MAGEFTH{_{\alpha\chi}}\PD{^\chi}\MAGEFTH{^{\alpha\beta}}
\nonumber\\ &\hspace{-30pt}
+\frac{3}{4}\PD{_\chi}\MAGEFTH{_{\alpha\beta}}\PD{^\chi}\MAGEFTH{^{\alpha\beta}}
+\dots\bigg].\label{EHExpansion}
\end{align}
By comparing~\cref{EHExpansion} with~\cref{ChiralD}, we see how the gauge-covariant kinetic operators both in chiral perturbation theory and in GR entail an infinite tower of corrections at higher orders in the field.
The first two terms in~\cref{EHExpansion} arise because the Riemannian curvature is linear in the metric perturbation at lowest order.
This observation will become important presently in the EFT of GR, and later on in the EFT of MAG.
For the moment, focussing only on the Einstein--Hilbert operator, these first two terms can be discarded as total derivatives that contribute only to the physics on the boundary.
The first physical terms in~\cref{EHExpansion} therefore turn out to be the marginal kinetic operators of gravity, which form the quadratic Lagrangian of massless Fierz--Pauli theory.
Classically, the higher-order corrections to~\cref{EHExpansion} constitute the non-linear completion of Fierz--Pauli theory under~$\DiffeomorphismGroup{}$ invariance with two derivatives, which is unique~\cite{Boulware:1974sr,Deser:1969wk,Deser:1987uk,Wald:1986bj,Butcher:2009ta}.
At the quantum level, however, the EFT of GR additionally continues at higher derivative order.
Taking the notation~$\big\{\GRG{n}{_i}\big\}$ for the set of couplings, the EFT of GR begins as
\begin{align} 
\mathcal{L} &= \sqrt{-g}\bigg[\GRG{4}{}\left(\Lambda^2\MAGR{}\right) + \frac{\GRG{6}{_1}}{\Lambda^2}\left(\Lambda^2\MAGR{_{\mu\nu}}\MAGR{^{\mu\nu}}\right) 
\nonumber\\ &\hspace{80pt}
+ \frac{\GRG{6}{_2}}{\Lambda^2}\left(\Lambda^2\MAGR{}^2\right) + \dots\bigg],\label{EFTGra}
\end{align}
where we use extra parentheses to indicate how powers of~$\Lambda$ must be absorbed to form each of the~$\big\{\OperatorO{i}{n}\big\}$.
As expected, the first operator in~\cref{EFTGra} is of the type~$\big\{\OperatorO{i}{4}\big\}$ given in~\cref{EHExpansion}; its uniqueness follows from Lovelock's theorem.\footnote{We neglect the cosmological constant, which is not relevant to the present discussion.}
Of course, the series in~\cref{EHExpansion} includes infinitely many non-linear~$\big\{\OperatorO{i}{n>4}\big\}$ operators at cubic and higher order in~$\MAGEFTH{_{\mu\nu}}$, but their coefficients are all determined by~$\GRG{4}{}$.
The fact that the first independent corrections are of the type~$\big\{\OperatorO{i}{6}\big\}$ rather than~$\big\{\OperatorO{i}{8}\big\}$ is due to the part of the Riemann curvature which is linear in~$\MAGEFTH{_{\mu\nu}}$.
Every operator must appear in~\cref{EFTGra} which is consistent with the guiding~$\DiffeomorphismGroup{}$ symmetry: for example, the possible third~$\big\{\OperatorO{i}{6}\big\}$ operator~$\MAGR{_{\mu\nu\sigma\rho}}\MAGR{^{\mu\nu\sigma\rho}}$ is only eliminated by the Gauss--Bonnet identity.
Unlike for the case of chiral perturbation theory, GR has the property that the~$\big\{\GRG{6}{_i}\big\}$ are needed at one-loop only if matter is present: the renormalisation of pure gravity begins at two-loops~\cite{tHooft:1974toh,tHooft:1973bhk,Deser:1974cy,Deser:1974nb,Goroff:1985th}. Note that the~$\big\{\OperatorO{i}{9}\big\}$ corrections, which are not shown in~\cref{EFTGra}, are inherently non-linear, being cubic in~$\MAGEFTH{_{\mu\nu}}$ even at lowest order: these do not contribute to the quadratic theory at all. We shall see in the case of MAG that inherently non-linear operators enter in already at marginal derivative order.

\paragraph*{Negative consequences} Both the pion model and GR clearly demonstrate that accessing the main virtues of the EFT approach fundamentally requires including all operators permitted by the defining symmetry. Arbitrarily selecting or omitting operators, on top of leading to inconsistencies when constructing a reliable perturbative expansion, undermines the central objective of EFT, namely, to achieve sufficient generality to fully capture the universal infrared (IR) behaviour of the candidate UV-completion.
The complete inclusion of all the permitted operators and the parallel expansion in~$E/\Lambda$ successfully tames another fatal characteristic of arbitrarily tuned modes: that of radiative instability. 
Quadratic operators with higher-derivative order, like the~$\big\{\OperatorO{i}{6}\big\}$ operators in~\cref{EFTGra}, are well known to introduce new ghost-like d.o.f whose mass is proportional to~$\Lambda$. Nevertheless, the stabilising symmetry automatically pushes such poles outside the region of perturbative validity of the defining EFT expansion in~$E/\Lambda$. As a result, the corresponding propagating d.o.f are understood to be heavy truncation artefacts~\cite{Weinberg:1995mt,Marzo:2024pyn}.
In the absence of a stabilising symmetry, new quadratic operators may be generated with the same number of derivatives as the tree-level starting Lagrangian: these propagate light states, which cannot be ignored.\footnote{Note that there have been some proposals for removing unwanted (Ostrogradsky-unstable) modes from gravitational theories at the quantum level~\cite{Mannheim:2011ds,Anselmi:2018ibi,Anselmi:2018tmf,Donoghue:2019fcb,Donoghue:2021eto}. See also~\cite{Glavan:2024cfs} for a novel approach to removing spurious d.o.f in the canonical framework.}
New light states are especially dangerous to MAG, whose rich tree-level spectrum is a fertile ground for ghosts and tachyons.
Without considering any quantum effects, it is already well known that the parameters of the generic MAG Lagrangian must be very finely tuned to prevent almost all of the d.o.f. in~\cref{DecompositionMAGA} from propagating.
Barring miraculous cancellations, it then seems somewhat likely that new physical d.o.f would tend to breach the finely-tuned unitarity of the tree-level Lagrangian~\cite{Marzo:2021iok,Marzo:2024pyn}.
To summarise, the very nature of metric-affine gravity actually makes the EFT approach more necessary than ever. Based on these considerations, we will next see in~\cref{Sec:Stab} how these principles can be applied to responsible model-building in the metric-affine context.

\begin{table*}[htbp]
	\includegraphics[width=\linewidth]{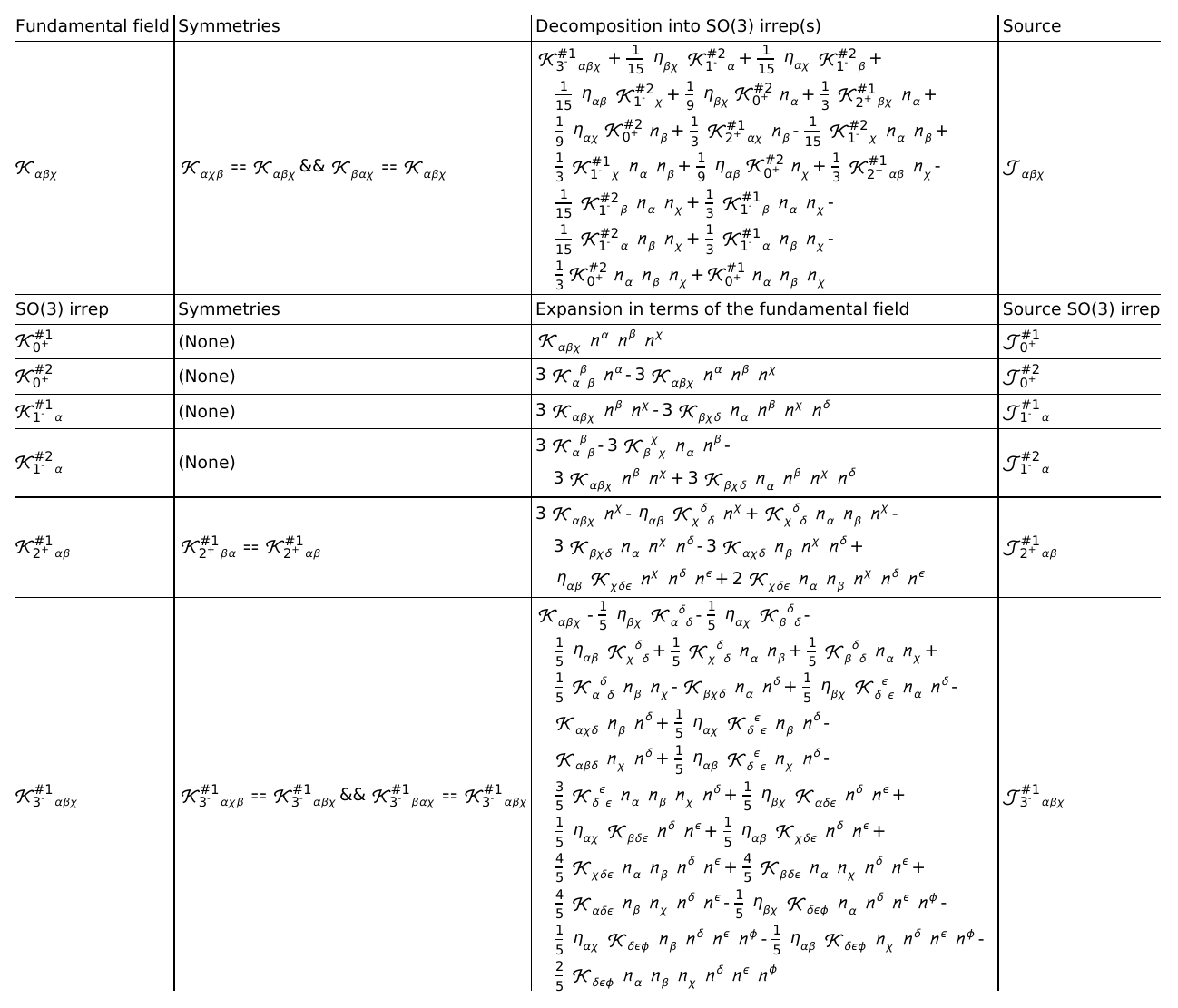}
	\caption{Output generated by \PSALTer{}. The field kinematics of the totally symmetric distortion defined in~\cref{SymmetricDistortion}. These definitions are used in~\cref{ParticleSpectrographM01,ParticleSpectrographm14,ParticleSpectrographM23,ParticleSpectrographm22,ParticleSpectrographm24,ParticleSpectrographm25}. Note that the collection of~$J^P_n$-states is substantially more limited than that given in~\cref{DecompositionMAGA} due to the condition of total symmetry. The unit-timelike vector is~$\tensor{n}{_{\mu}}\equiv\tensor{k}{_{\mu}}/\sqrt{\tensor{k}{^\nu}\tensor{k}{_\nu}}$, where~$\tensor{k}{_\mu}$ is the massive particle four-momentum.}
\label{FieldKinematicskNS}
\end{table*}

\section{The algorithm}\label{Sec:Stab}

\paragraph*{General Lagrangian} In~\cref{Sec:Pred} we emphasised how the small collection of funamental theories that are actually realised in nature (as opposed to those which have only been proposed or speculated about) all have their place as EFTs. The EFT status applies even to GR, and we will now see whether it can be applied also to MAG. The field reparameterisation in~\cref{postriem} shifts the \emph{affine} or \emph{Palatini} basis~$\{\MAGg{_{\mu\nu}}, \MAGA{_\mu^\rho_\nu}\}$ into the \emph{post-Riemann} basis~$\{ \MAGg{_{\mu\nu}} , \Dis{_\mu^\rho_\nu}\}$.
As emphasised in~\cref{Sec:Intro}, this change of variables is fully invertible, and thus has no impact on the physical content of the theory.
In the post-Riemannian variables, the quantities defined in~\cref{MAGFDef,MAGTDef,MAGQDef} are redefined
\begin{subequations}
\begin{align}
	\MAGF{_{\mu\nu}^\rho_\sigma} &\equiv \MAGR{_{\mu\nu}^\rho_\sigma} + 2\big(\tensor{\partial}{_{[\mu}}\Dis{_{\nu]}^\rho_\sigma}+\Con{_{[\mu|}^\rho_\alpha}\Dis{_{|\nu]}^\alpha_\sigma} +  
	\nonumber\\&\hspace{15pt}
	+ \Dis{_{[\mu|}^\rho_\alpha}\Con{_{|\nu]}^\alpha_\sigma} +\Dis{_{[\mu|}^\rho_\alpha}\Dis{_{|\nu]}^\alpha_\sigma} \big) \,, 
	\label{MAGFDecomposition}\\
    \MAGT{_\mu^\alpha_\nu} &\equiv 2\Dis{_{[\mu|}^\alpha_{|\nu]}} \,, 
	\label{MAGTDecomposition}\\ 
    \MAGQ{_{\lambda\mu\nu}} &\equiv 2\Dis{_\lambda^\alpha_{(\mu|}}\MAGg{_{\alpha|\nu)}} \,, \label{MAGQDecomposition}
\end{align}
\end{subequations}
where in~\cref{MAGFDecomposition} the standard Riemann curvature tensor was introduced in~\cref{MAGRDef}.
The kinematic restriction in~\cref{SymmetricDistortion} evidently eliminates~\cref{MAGTDecomposition}.
With this restriction, the derivative expansion of the most general parity-preserving Lagrangian, restricted only by~$\DiffeomorphismGroup{}$ invariance and with the notation~$\big\{\MAGG{n}{_i}\big\}$ for the couplings, is given by
\begin{align}
	\mathcal{L} & =  \sqrt{-g}\bigg[\MAGG{2}{_1}\Lambda^2\Dis{_{\alpha\mu\nu}} \Dis{^{\alpha\mu\nu}} + \MAGG{2}{_2}\Lambda^2 \Dis{^{\alpha}_{\alpha}^{\beta}} \Dis{_{\beta}^{\chi}_{\chi}}
	\nonumber\\& \hspace{5pt}
	+ \MAGG{4}{_1} \CD{_{\beta}}\Dis{_{\chi}^{\delta}_{\delta}} \CD{^{\chi}}\Dis{^{\alpha}_{\alpha}^{\beta}} + \MAGG{4}{_2} \CD{_{\chi}}\Dis{_{\beta}^{\delta}_{\delta}} \CD{^{\chi}}\Dis{^{\alpha}_{\alpha}^{\beta}}
	\nonumber\\& \hspace{5pt}
	+ \MAGG{4}{_3} \CD{_{\alpha}}\Dis{^{\alpha\beta\chi}} \CD{_{\delta}}\Dis{_{\beta\chi}^{\delta}} + \MAGG{4}{_4} \CD{^{\chi}}\Dis{^{\alpha}_{\alpha}^{\beta}} \CD{_{\delta}}\Dis{_{\beta\chi}^{\delta}} 
	\nonumber\\& \hspace{5pt}
	+ \MAGG{4}{_7} \CD{_{\delta}}\Dis{_{\alpha\beta\chi}} \CD{^{\delta}}\Dis{^{\alpha\beta\chi}} 
	+ \MAGG{4}{_8} \left(\Lambda^2\MAGR{}\right)
	\nonumber\\& \hspace{5pt}
	+ \frac{\MAGG{5}{_1}}{\Lambda} \left(\Lambda\MAGR{^{\mu\nu}}\CD{_\nu}\Dis{_\mu^\rho_\rho}\right)
	+ \frac{\MAGG{5}{_2}}{\Lambda} \left(\Lambda\MAGR{}\CD{_\nu}\Dis{^\mu_\mu^\nu}\right)
	\nonumber\\& \hspace{5pt}
	+ \frac{\MAGG{6}{_{1}}}{\Lambda^2} \left(\Lambda^2\MAGR{_{\mu\nu}}\MAGR{^{\mu\nu}}\right)
	+ \frac{\MAGG{6}{_{2}}}{\Lambda^2} \left(\Lambda^2\MAGR{}^2\right)
	+\dots\bigg],\label{FlatLag}
\end{align}
with~$\nabla_\mu$ being the covariant derivative constructed from the Levi--Civita connection~$\Con{_\mu^\nu_\rho}$, and where we notice the obvious parallels between~\cref{FlatLag,EFTGra}.
As with~\cref{EFTGra}, we use extra parentheses to indicate how powers of~$\Lambda$ must be absorbed to form each~$\big\{\OperatorO{i}{n}\big\}$.
Moreover, we also omit from~\cref{FlatLag} a somewhat large collection of possible lower-derivative operators which are inherently non-linear in the perturbed fields, such as the~$\big\{\OperatorO{i}{4}\big\}$ operators which are quartic in~$\Dis{_\mu^\rho_\nu}$ and which cannot yet be excluded on any physical grounds.
The point is that these operators are not part of the quadratic sector of MAG.
This quadratic sector must be healthy \emph{without} non-linear effects if MAG is to be a perturbative QFT, and quadratic health is already a big enough challenge for standalone investigation.
As with the EFT of GR,~\cref{FlatLag} includes higher-derivative operators which are quadratic in the curvature. Recall that, in the GR case, these operators generally bring in new ghost-like d.o.f., but that the EFT interpretation allows us to discard these as heavy truncation artefacts.
The format of~\cref{FlatLag} splendidly illustrates our point from~\cref{Sec:Intro} that the~$\DiffeomorphismGroup{}$-type version of MAG (the common interpretation in the modern literature) is really just a theory of matter coupled (non-)minimally to Riemannian gravity, as we shall now see.

\paragraph*{Choice of Lagrangian} By focusing on the quadratic sector of MAG as in~\cref{FlatLag}, we observe that the only quadratic mixing between~$\MAGEFTH{_{\mu\nu}}$ and~$\Dis{_\mu^\rho_\nu}$ arises through the couplings~$\big\{\MAGG{5}{_i}\big\}$. In EFT terms, this indicates that such mixing first emerges at mass dimension five and at the level of three-derivative operators, making it subdominant compared to the allowed two-derivative, dimension-four operators.
This aspect is most apparent when the gravitational fields are expressed (as here) in their physically meaningful engineering dimensions using~\cref{MAGEFTHDef} such that, schematically,
\begin{subequations}
\begin{gather}
	\Lambda^2\MAGR{}\sim\MAGEFTH{}\partial^2\MAGEFTH{}\sim\OperatorO{i}{4},
	\\
	\Lambda\MAGR{}\nabla\Dis{}\sim\partial^2\MAGEFTH{}\partial\Dis{}\sim\OperatorO{i}{5},
	\\
	\Lambda^2\MAGR{}^2\sim\partial^2\MAGEFTH{}\partial^2\MAGEFTH{}\sim\OperatorO{i}{6}.
\end{gather}
\end{subequations}
The fact that~$\MAGEFTH{_{\mu\nu}}$ is of mass-dimension one is not a convention: it is a physical necessity following from its identity as a propagating bosonic field.
The EFT framework therefore justifies, at the dominant quadratic order in perturbed fields, a decoupled analysis limited to the distortion terms in~\cref{FlatLag}.
In other words, the gravitational sector completely falls away, and we are left with the pure matter theory.
This deals with the gravitational dynamics; to deal with the background we select Minkowski spacetime, as it is the background relevant to high-energy scattering. Regardless of energy scales, unitary and causal propagation on flat spacetime is likely to be a necessary condition for physicality.\footnote{Similarly, acausal propagation in an electromagnetic background disposes of theories with healthy Minkowskian propagation~\cite{Velo:1969bt}. Whilst we put forward the Minkowski background as being the most relevant, we note that there is a growing literature on alternatives, such as (anti-)de Sitter. See for example~\cite{Annala:2022gtl} and references therein for applications in metric-affine theory, and also the series in~\cite{Karananas:2024hoh,Karananas:2024qrz}.} Consequently, without loss of generality, our analysis will focus on the flat limit of~\cref{FlatLag}, specifically within the decoupled sector involving the distortion tensor
\begin{align}
	\mathcal{L} & =  \MAGG{2}{_1}\Lambda^2\Dis{_{\alpha\mu\nu}} \Dis{^{\alpha\mu\nu}} + \MAGG{2}{_2}\Lambda^2 \Dis{^{\alpha}_{\alpha}^{\beta}} \Dis{_{\beta}^{\chi}_{\chi}} 
	\nonumber\\& \hspace{5pt}
	+ \MAGG{4}{_1} \PD{_{\beta}}\Dis{_{\chi}^{\delta}_{\delta}} \PD{^{\chi}}\Dis{^{\alpha}_{\alpha}^{\beta}} + \MAGG{4}{_2} \PD{_{\chi}}\Dis{_{\beta}^{\delta}_{\delta}} \PD{^{\chi}}\Dis{^{\alpha}_{\alpha}^{\beta}}
	\nonumber\\& \hspace{5pt}
	+ \MAGG{4}{_3} \PD{_{\alpha}}\Dis{^{\alpha\beta\chi}} \PD{_{\delta}}\Dis{_{\beta\chi}^{\delta}} + \MAGG{4}{_4} \PD{^{\chi}}\Dis{^{\alpha}_{\alpha}^{\beta}} \PD{_{\delta}}\Dis{_{\beta\chi}^{\delta}} 
	\nonumber\\& \hspace{5pt}
	+ \MAGG{4}{_7} \PD{_{\delta}}\Dis{_{\alpha\beta\chi}} \PD{^{\delta}}\Dis{^{\alpha\beta\chi}} 
	\, . \label{VeryFlatLag}
\end{align}
The quadratic theory in~\cref{VeryFlatLag}, set up on Minkowski spacetime, is essentially analogous to the general quadratic ansatz that one can set up in~$\MAGEFTH{_{\mu\nu}}$. From the~$\MAGEFTH{_{\mu\nu}}$-ansatz, the tuned Fierz--Pauli theory in~\cref{EHExpansion} emerges as the unique two-derivative~$\DiffeomorphismGroup{}$-invariant model, from which one can bootstrap the whole non-linear Einstein--Hilbert Lagrangian~\cite{Boulware:1974sr,Deser:1969wk,Deser:1987uk,Wald:1986bj,Butcher:2009ta}. We want to look for similar symmetric foundations within~\cref{VeryFlatLag}.

\paragraph*{Non-linear symmetries} We now consider the general relationship between interactions and symmetries~\cite{Deser:1963zzc,Barnich:1993vg,Berends:1984rq,Boulanger:2000rq,Buchbinder:2021rfm,Buchbinder:2021rmy} in order to highlight the pivotal, defining role played by the quadratic sector. We emphasise that the following structure is general, and independent of the specific non-linear realisation that completes~\cref{VeryFlatLag}. 
The action~$\mathcal{S}\equiv\int\mathrm{d}^4x \,\mathcal{L}$ corresponding to the full non-linear completion of the Lagrangian~\cref{VeryFlatLag}, can be obtained as a series by introducing a perturbative parameter~$\varepsilon$~\cite{Barnich:1993vg} (see~\cite{Butcher:2009ta} for this formalism in the gravity sector)
\begin{equation}\label{eq:seriesS}
\begin{gathered}
	\mathcal{S}\left[K\right]=\sum_{n=2}^\infty \varepsilon^n\mathcal{S}_{(n)}\left[\mathcal{K}\right],
	\\
	\mathcal{S}_{(n)}\left[\mathcal{K}\right]\equiv\frac{1}{n!}\bigg[\frac{\PD{}^n}{\PD{}\varepsilon^n}\mathcal{S}\left[\varepsilon\mathcal{K}\right]\bigg]_{\varepsilon=0}, \quad \tensor{\mathcal{K}}{_\mu^\alpha_\nu}\equiv\frac{1}{\varepsilon}\Dis{_\mu^\alpha_\nu},
\end{gathered}
\end{equation}
where~$\mathcal{S}_{(n)}\left[\mathcal{K}\right]$ is exclusively~$n$-th order in~$\tensor{\mathcal{K}}{_\mu^\alpha_\nu}$. The fact that the series in~\cref{eq:seriesS} starts with the quadratic part is a consequence of the Minkowski~$\Dis{_{\alpha\beta\chi}}=0$ background being a solution of the equations of motion. Similarly, the non-linear completion of the gauge transformation takes the form
\begin{equation}\label{eq:seriesdelta}
	\begin{gathered}
		\tensor{\delta(K)}{_{\alpha\beta\chi}}=\sum_{n=0}^\infty \varepsilon^n\tensor{\delta_{(n)}(\mathcal{K})}{_{\alpha\beta\chi}},\\
	\tensor{\delta_{(n)}(\mathcal{K})}{_{\alpha\beta\chi}}\equiv 
	\frac{1}{n!}\bigg[\frac{\PD{}^n}{\PD{}\varepsilon^n}\tensor{\delta(\varepsilon\mathcal{K})}{_{\alpha\beta\chi}}\bigg]_{\varepsilon=0}.
	\end{gathered}
\end{equation}
When~\cref{eq:seriesdelta} is substituted into~\cref{eq:seriesS}, the symmetry is realised order-by-order in an infinite perturbative hierarchy of exact equations
\begin{subequations}
\begin{gather}
	\delta_{(0)}\left(\mathcal{S}_{(2)}\right)=0,\label{eq:syms1}\\
	\delta_{(1)}\left(\mathcal{S}_{(2)}\right)+\delta_{(0)}\left(\mathcal{S}_{(3)}\right)=0,\label{eq:syms2}\\
	\delta_{(2)}\left(\mathcal{S}_{(2)}\right)+\delta_{(1)}\left(\mathcal{S}_{(3)}\right)+\delta_{(0)}\left(\mathcal{S}_{(4)}\right)=0,\label{eq:syms3}\\
	\vdots\nonumber\\
	\sum_{m=0}^{n}\delta_{(m)}\left(\mathcal{S}_{(n+2-m)}\right)=0,\quad n\geq 2.\label{eq:syms4}
\end{gather}
\end{subequations}
The first equation in~\cref{eq:syms1} is an identity, not a constraint: it is the foundation of the non-linear completion of the gauge symmetry -- assuming that such a non-linear completion exists at all~\cite{Boulanger:2000rq,Bekaert:2010hp}. Accordingly, we will focus on identifying such exact foundations at quadratic order.

\paragraph*{Un-free symmetries} As illustrated in~\cref{DecompositionMAGA}, a rank three field contains many independent~$J^P_n$ states. It is in principle possible to introduce a symmetry corresponding to the removal of an individual~$J^P_n$ sector, independently of all the others. This, however, has certain consequences. Given a Lorentz-covariant generator field, the process of extracting specific~$J^P_n$ components requires the use of the orthonormal projectors
\begin{equation}\label{eq:proj}
	\tensor*{\Theta}{_\mu^\nu}\equiv\Kronecker{_\mu^\nu}-\frac{\PD{_\mu}\PD{^\nu}}{\PD{_\alpha}\PD{^\alpha}},\quad
	\tensor*{\Omega}{_\mu^\nu}\equiv\frac{\PD{_\mu}\PD{^\nu}}{\PD{_\alpha}\PD{^\alpha}},
\end{equation}
respectively for the transverse and longitudinal parts. Notice how the projectors in~\cref{eq:proj} are both inherently non-local: to restore locality in a~$J^P_n$-projector, i.e. a polynomial in the~$\tensor*{\Theta}{_\mu^\nu}$ and~$\tensor*{\Omega}{_\mu^\nu}$, one must multiply by some adequate power of~$\PD{^\alpha}\PD{_\alpha}$. The consequence of this procedure is a transformation law featuring multiple derivatives acting on the unconstrained generator field. In general, each higher-derivative gauge symmetry written in terms of an unconstrained generator can be written in terms of a lower-derivative gauge symmetry acting on a constrained generator~\cite{Kaparulin:2019quz,Abakumova:2022esw,Abakumova:2023anf,Bittencourt:2025roa,Francia:2013sca,Campoleoni:2012th}. Constrained or un-free symmetries are ubiquitous in higher-spin theories, and are an active area of research. A prominent example is unimodular gravity, which features a reduced symmetry compared to the~$\DiffeomorphismGroup{}$ invariance of GR~\cite{Alvarez:2023utn,Barvinsky:2022guw,Jirousek:2020vhy,Percacci:2017fsy,Alvarez:2006uu}. Specifically, it is invariant only under volume-preserving coordinate transformations, generated by transverse diffeomorphisms such that~\cref{DiffMAGH} is replaced by
\begin{equation}
\tensor{\delta(h)}{_{\mu\nu}}=\PD{_{\mu}}\GenT{_{\nu}}+\PD{_{\nu}}\GenT{_{\mu}}+\mathcal{O}(h),\label{TransMAGH} \end{equation}
where the transverse generator~$\GenT{_{\mu}}$ is defined in terms of the unconstrained generator~$\Gen{_\nu}$ as
\begin{equation}\label{TransGen}
	\GenT{_{\mu}} \equiv\tensor*{\Theta}{_\mu^\nu}\Gen{_\nu} \implies
	\PD{^\mu}\GenT{_\mu}=0,
\end{equation}
such that~\cref{TransGen} implies a constraint. The symmetry in~\cref{TransMAGH} can be reformulated in terms of an unconstrained generator with a minimal power of extra derivatives, for instance
\begin{eqnarray}
	\GenT{_{\mu}} \equiv \PD{^{\nu}} \Gen{_{[\mu \nu]}} \, . \nn
\end{eqnarray}
Unlike their unconstrained counterparts, constrained gauge symmetries necessitate a significantly more involved quantization procedure and recursive gauge-fixing process, giving rise to the `\emph{ghosts of ghosts}' scenario characteristic of reducible symmetries~\cite{Siegel:1980jj,Batalin:1983pz,Batalin:1983wj,Ferraro:1992ek}. Additionally, the modified Noether identity introduces several field-dependent on-shell constants, whose dynamical interpretation remains unclear~\cite{Abakumova:2022esw}.

\paragraph*{Choice of symmetries} Following from the considerations above, we admit three possible kinds of free, linear symmetries
\begin{subequations}
\begin{align}
\tensor{\delta_{(0)}(K)}{_{\alpha\beta\chi}}&=
t_{2}^{}\PD{_{\beta}}\PD{_{\alpha}}\PD{_{\chi}}\xi
\nonumber\\&\
+r_{1}^{}\left(\G{_{\chi\beta}}\PD{_{\alpha}}\xi+\G{_{\chi\alpha}}\PD{_{\beta}}\xi+\G{_{\alpha\beta}}\PD{_{\chi}}\xi\right)
,\label{eq:syms0}\\
\tensor{\delta_{(0)}(K)}{_{\alpha\beta\chi}}&=
t_{2}^{}\left(\G{_{\chi\beta}}\PD{_{\delta}}\PD{_{\alpha}}\tensor{\xi}{^{\delta}}+\G{_{\chi\alpha}}\PD{_{\delta}}\PD{_{\beta}}\tensor{\xi}{^{\delta}}+\G{_{\alpha\beta}}\PD{_{\delta}}\PD{_{\chi}}\tensor{\xi}{^{\delta}}\right)
\nonumber\\&\
+t_{3}^{}\left(\PD{_{\alpha}}\PD{_{\chi}}\tensor{\xi}{_{\beta}}+\PD{_{\beta}}\PD{_{\alpha}}\tensor{\xi}{_{\chi}}+\PD{_{\beta}}\PD{_{\chi}}\tensor{\xi}{_{\alpha}}\right)
\nonumber\\&\ \ \ 
+r_{1}^{}\left(\tensor{\xi}{_{\chi}}\G{_{\alpha\beta}}+\tensor{\xi}{_{\beta}}\G{_{\chi\alpha}}+\tensor{\xi}{_{\alpha}}\G{_{\chi\beta}}\right)
,\label{eq:syms01}\\
\tensor{\delta_{(0)}(K)}{_{\alpha\beta\chi}}&=
r_{1}^{}\left(\G{_{\chi\beta}}\PD{_{\delta}}\tensor{\xi}{_\alpha^\delta}+\G{_{\chi\alpha}}\PD{_{\delta}}\tensor{\xi}{_\beta^\delta}+\G{_{\alpha\beta}}\PD{_\delta}\tensor{\xi}{_\chi^\delta}\right)
\nonumber\\&\
+r_{2}^{}\left(\G{_{\chi\beta}}\PD{_{\alpha}}\tensor{\xi}{^{\delta}_\delta}+\G{_{\chi\alpha}}\PD{_{\beta}}\tensor{\xi}{^{\delta}_\delta}+\G{_{\alpha\beta}}\PD{_{\chi}}\tensor{\xi}{^\delta_\delta}\right)
\nonumber\\&\
+r_{3}^{}\left(\PD{_{\alpha}}\tensor{\xi}{_{\chi\beta}}+\PD{_{\beta}}\tensor{\xi}{_{\chi\alpha}}+\PD{_{\chi}}\tensor{\xi}{_{\alpha\beta}}\right)
,\label{eq:syms02}
\end{align}
\end{subequations}
where~\cref{eq:syms0,eq:syms01,eq:syms02} are generated by a single-d.o.f scalar~$\Gen{}$, a four-d.o.f vector~$\Gen{_\mu}$ and a symmetric rank-two tensor~$\Gen{_{\mu\nu}}$ with 10 d.o.f, respectively. Together with~\cref{VeryFlatLag}, these transformation ans\"atze are substituted into the identity in~\cref{eq:syms1}. This results in a simultaneous system of bilinear equations in the symmetry parameters and Lagrangian couplings. This system is reduced by standard algebraic means so as to identify viable quantum foundations for MAG.

\paragraph*{Particle spectroscopy} For each of the models identified by this algorithm, the full spectrum of tree-level particles is determined. This is done by computing the propagator of the model as follows. A collection of test sources~$\tensor{\mathcal{J}}{^{\mu\nu\sigma}}$ is used to probe the quadratic theory in~\cref{eq:seriesS} according to
\begin{equation}\label{GenLag}
	\mathcal{S}_{(2)}=\int\mathrm{d}^4x\ \Dis{_{\mu\nu\sigma}}\left[\tensor{\mathcal{O}}{^{\mu\nu\sigma}_{\alpha\beta\gamma}}\Dis{^{\alpha\beta\gamma}}-\tensor{\mathcal{J}}{^{\mu\nu\sigma}}\right],
\end{equation}
where the operator~$\tensor{\mathcal{O}(\partial)}{^{\mu\nu\sigma}_{\alpha\beta\gamma}}$ is the linearized wave operator of the theory: a tensor polynomial of up to quadratic order in~$\PD{_\mu}$, and otherwise linearly parameterised by the Lagrangian coupling coefficients of varying mass dimension. The saturated propagator associated with~\cref{GenLag} is defined formally by transitioning to momentum space via~$\PD{_\mu}\mapsto -i\tensor{k}{_\mu}$ and `inverting' the wave operator, before sandwiching between sources, i.e.,
\begin{equation}\label{Propagator}
	\Pi(k)\equiv\tensor{\mathcal{J}^\dagger(k)}{_{\mu\nu\sigma}}\tensor{\mathcal{O}^{-1}(k)}{^{\mu\nu\sigma}_{\alpha\beta\gamma}}\tensor{\mathcal{J}(k)}{^{\alpha\beta\gamma}}.
\end{equation}
In cases where the theory has some gauge symmetry, this `inverse' may not exist due to degeneracies in the wave operator. The symmetries that cause these degeneracies, however, also impose constraints on the source currents. When the `inverse' wave operator is sandwiched between the sources as in~\cref{Propagator}, its infinite sectors are eliminated by these constraints. In~$k$-space, the positions of the poles of~$\Pi(k)$ determine the masses of the physical states, with real masses corresponding to non-tachyonic particles. The residue of~$\Pi(k)$ at each pole must be positive definite, otherwise the state is a ghost. For massive states, a~$J^P_n$-basis analogous to that shown in~\cref{DecompositionMAGA,DecompositionMAGH} is well suited for reducing the wave operator inversion to a linear algebra problem. For massless states, spin is superseded by helicity, and the calculation has to be done component-wise. This process always determines the massless polarisations and their no-ghost conditions, but the associated values of~$J$ and~$P$ are not always self-evident for massless states. In this paper we use the \PSALTer{} software package to perform the spectroscopy~\cite{Barker:2024juc}. As shown in~\cref{FieldKinematicskNS}, the software can accommodate a structured tensor field such as that defined in~\cref{SymmetricDistortion}, and automatically determines its constituent~$J^P_n$ states.

\section{Results}\label{Results}

\begin{figure*}[htbp]
	\includegraphics[width=\linewidth]{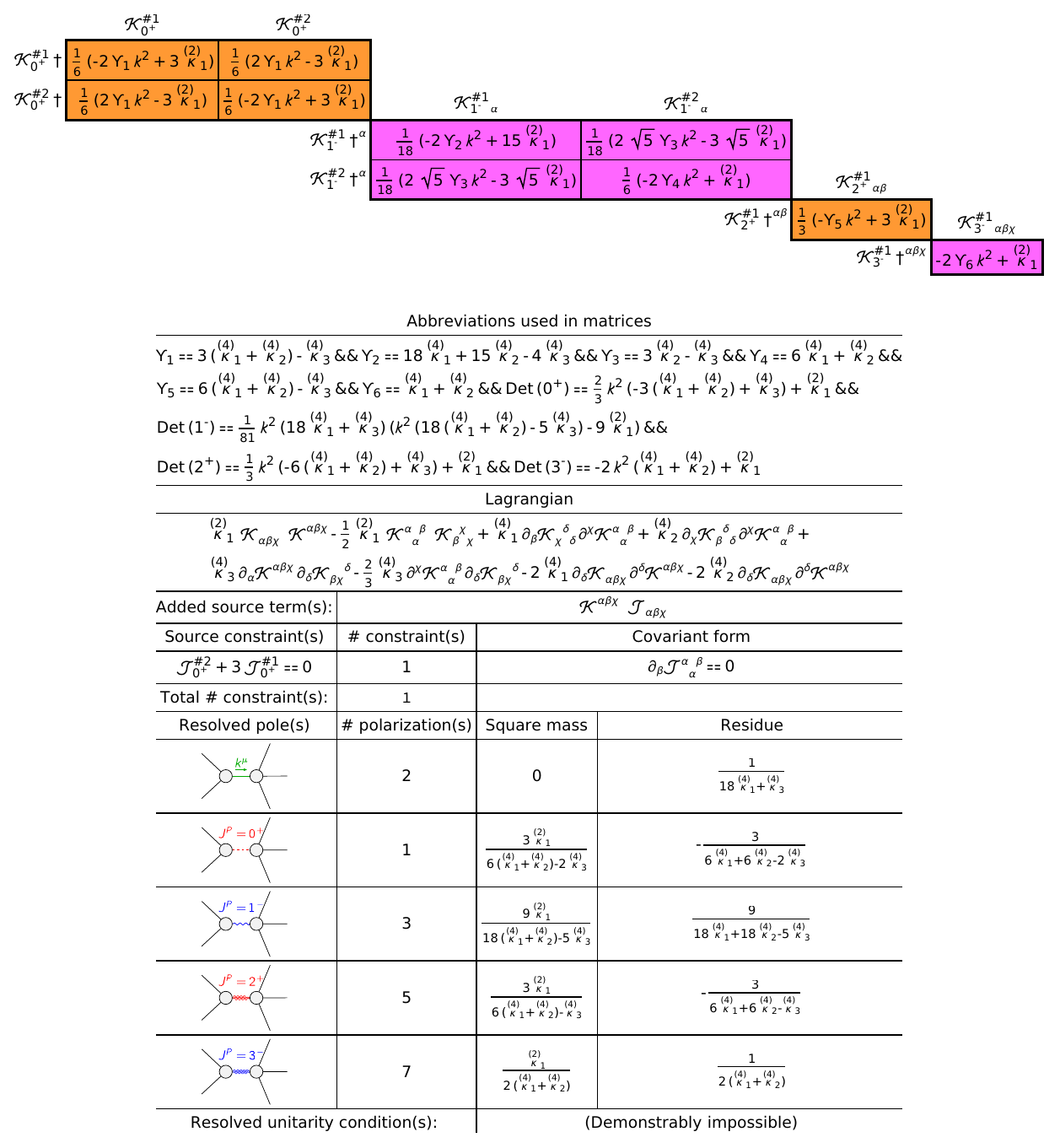}
	\caption{Output generated by \PSALTer{}. The spectrograph of~\Model{0}{1}, as defined in~\cref{M0}. All notation is defined in~\cref{FieldKinematicskNS}.}
\label{ParticleSpectrographM01}
\end{figure*}

\paragraph*{Scalar generator} The symmetry is given in~\cref{eq:syms0}, and the results of the survey are shown in~\cref{M0}. One model presents massive states with a typical multi-sector propagation showing conflicting requests for stability (see~\cref{ParticleSpectrographM01}). The remaining two models propagate massless states with higher-order poles. We find that, in our parity-preserving, totally symmetric distortion framework, no metric-affine model can exist which relies on the single scalar generator to stabilise a ghost-free spectrum against the introduction of consistent interactions.

\begin{table*}[htbp]
	\caption{\label{M0} Models with scalar generators. Symmetries are to be applied to~\cref{eq:syms0}, and the model constraints are to be applied to~\cref{VeryFlatLag}. The spectrum is inconsistent (\SampleInconsistent{}), consistent (\SampleConsistent{}), empty (\SampleEmpty{}) or leads to further analysis (\SampleMore{}). Resolved poles may be massive ($\MassiveParticle{J}{P}$), massless ($\MasslessParticle{J}{P}$) or include multiple massless states of unspecified~$J^P$~($\UnknownMasslessParticle{}$).}
\begin{center}
\renewcommand{\arraystretch}{1.7}%
	\begin{tabularx}{\linewidth}{c|p{2cm}|X|c|c}
\hline\hline
	\# & Symmetry & Lagrangian & Unitarity & Spectrum \\
\hline
	\Model{0}{1} &~$t_2=0$ &~$\big(\MAGG{4}{_4} +\tfrac{2}{3} \MAGG{4}{_3}=0 \big)\wedge\big(  \MAGG{4}{_7} +2 (\MAGG{4}{_1} + \MAGG{4}{_2})=0 \big)\wedge\big(  \MAGG{2}{_2}+\tfrac{1}{2}\MAGG{2}{_1}\big)$ & \Inconsistent{} &~$\UnknownMasslessParticle{}\wedge\MassiveParticle{0}{+}\wedge\MassiveParticle{1}{-}\wedge\MassiveParticle{2}{+}\wedge\MassiveParticle{3}{-}$ \\
	\Model{0}{2} &~$r_1=0$ &~$\big(\MAGG{4}{_4} + 2 (\MAGG{4}{_1} + \MAGG{4}{_2})=0 \big)\wedge\big(  \MAGG{4}{_7}- \MAGG{4}{_1} - \MAGG{4}{_2} + \MAGG{4}{_3}=0 \big)\wedge\big(  \MAGG{2}{_1} - \MAGG{2}{_2}=0\big)$ & \Inconsistent{} &~$\UnknownMasslessParticle{}$ \\
	\Model{0}{3} &~$\big(r_1 \neq 0 \big)\wedge\big( t_2 \neq 0\big)$ &~$\big(\MAGG{4}{_3} = 3 (\MAGG{4}{_1} + \MAGG{4}{_2}) \big)\wedge\big( \MAGG{4}{_4} = \MAGG{4}{_7} \big) \wedge \big( \MAGG{4}{_7} = -2 (\MAGG{4}{_1} + \MAGG{4}{_2}) \big)\wedge\big(  \MAGG{2}{_1} = \MAGG{2}{_2}\big)$ & \Inconsistent{} &~$\UnknownMasslessParticle{}$ \\
\hline\hline
\end{tabularx}
\end{center}
\end{table*}

\begin{figure*}[htbp]
	\includegraphics[width=\linewidth]{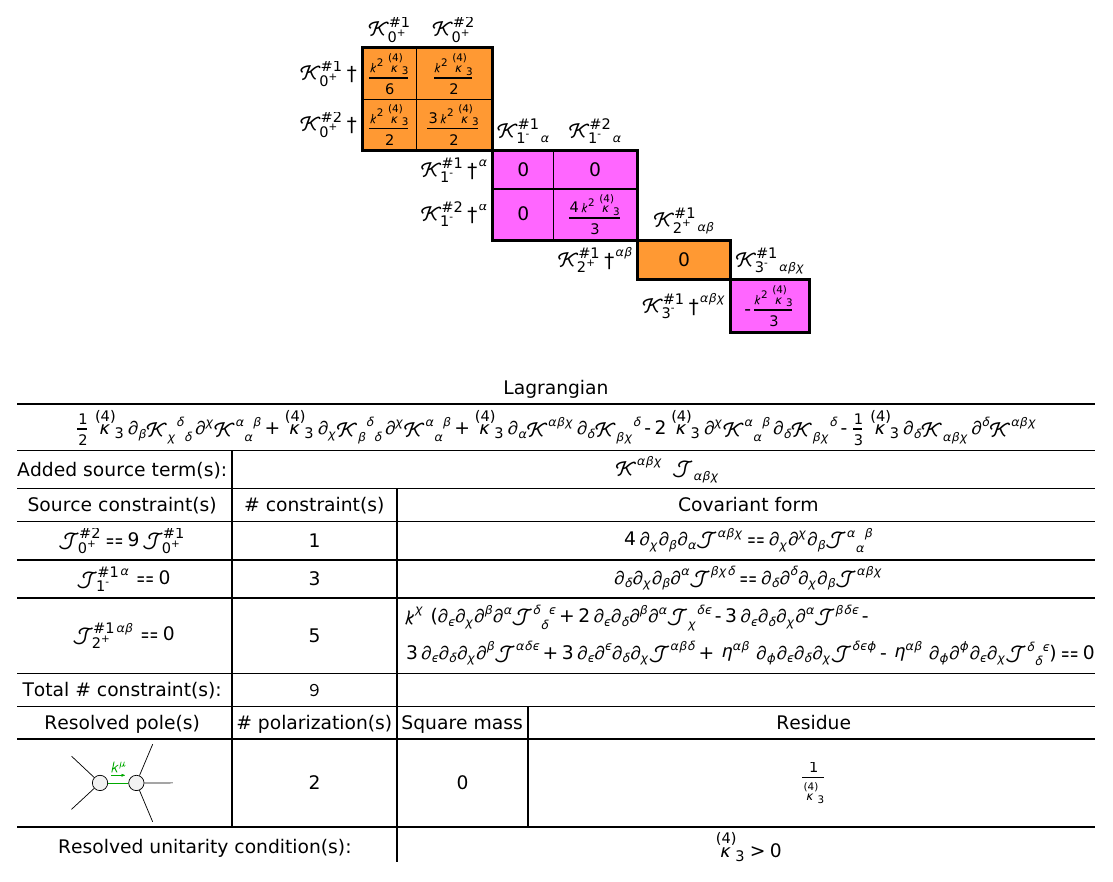}
	\caption{Output generated by \PSALTer{}. The spectrograph of~\SubModel{1}{4}, as defined in~\cref{m1}. All notation is defined in~\cref{FieldKinematicskNS}.}
\label{ParticleSpectrographm14}
\end{figure*}

\paragraph*{Vector generator} The symmetry is given in~\cref{eq:syms01}, and the results of the survey are shown in~\cref{M1}.
Note that symmetry transformations generated by a local vector~$\Gen{_{\mu}}$ are usually referred to as \emph{projective} transformations; they represent a natural candidate for the consistent removal of unwanted states within~$\Dis{_{\mu \nu \rho}}$~\cite{Baikov:1992uh,Barker:2024dhb,Jimenez-Cano:2022sds,Garcia-Parrado:2020lpt,BeltranJimenez:2020sqf}.  
Apart from the various inconsistent massive and purely massless models, a third possibility is presented by
model~\Model{1}{2}. The corresponding Lagrangian is
\begin{align}\label{eq:rank1L}
	\mathcal{L} = \MAGG{4}{_1}  \PD{^{\chi}}\Dis{_\chi^\delta_\delta}\PD{_\beta}\Dis{^\alpha_\alpha^\beta} \, ,
\end{align}
which would na\"ively appear to propagate the longitudinal component of the vector~$\Dis{_\chi^\delta_\delta}$.
In reality, however,~\cref{eq:rank1L} does not propagate any state unless a mass term is added, as is already known from the corresponding vector theory.
Elsewhere, the symmetry of~\Model{1}{5} varies smoothly with two parameters, whilst the definition of the model itself exhibits a dependence on these parameters: we term such symmetries \emph{parametric}.
Applying the spectral analysis directly to a model with parametric symmetry, without careful examination, risks missing distinct cases that are not smoothly connected to the generic model. This issue arises from the discontinuous dependence of propagation on the parameters: certain configurations of the free parameters may be finely tuned to generate additional symmetries.
First, we observe that the singular points of the given solution~$z_1=0$ or~$z_2=0$ are indeed already covered by the general scan, and correspond to the~\Model{1}{2} and~\Model{1}{4} cases.
Additional branches can be identified through a direct analysis of the~$J^P_n$ sectors.
The corresponding matrices reveal five opportunities for symmetry enhancement by a tuning of the~$\big\{z_i\big\}$. Of these, the case~$z_1=z_2$ corresponds to solution~\Model{1}{3}.
Together with the generic solution~\Model{1}{5}, the four new branches are presented in~\cref{m1}.
At the level of the particle content, we identify one case that successfully stabilizes the ghost-free spectrum, in the model~\SubModel{1}{4}, which enhances the symmetry of~\Model{1}{5} by eliminating the spin-two sector through the condition~$z_1=2z_2$ (see~\cref{ParticleSpectrographm14}). For completeness, we quote the full form of the stabilising symmetry, which is
\begin{align}
	\tensor{\delta_{(0)}(K)}{_{\alpha\beta\chi}} & =
	 \frac{1}{4} \left(\G{_{\chi\beta}} \PD{_{\delta}}\PD{_{\alpha}}\Gen{^{\delta}} + \G{_{\chi\alpha}} \PD{_{\delta}}\PD{_{\beta}} \Gen{^{\delta}} + \G{_{\alpha\beta}} \PD{_{\delta}}\PD{_{\chi}} \Gen{^{\delta}}\right)
	 \nonumber\\& \hspace{15pt}
	+\PD{_{\alpha}}\PD{_{\chi}}\Gen{_{\beta}} + \PD{_{\beta}}\PD{_{\alpha}}\Gen{_{\chi}} + \PD{_{\beta}}\PD{_{\chi}} \Gen{_{\alpha}}.\label{eq:rank1sym}
\end{align}    
It comes as a surprising result that the corresponding Lagrangian 
\begin{align} \label{SurpriseFron}
	\mathcal{L} & = \MAGG{4}{_3} \bigg[
		\PD{_\beta} \Dis{_\chi^\delta_\delta}\PD{^\chi}\Dis{^\alpha_\alpha^\beta}+ 2\PD{_\chi} \Dis{_\beta^\delta_\delta}\PD{^\chi}\Dis{^\alpha_\alpha^\beta}
		\nonumber\\& \hspace{30pt}
	+ 2 \PD{_\alpha}\Dis{^{\alpha\beta\chi}}\PD{_\delta}\Dis{_{\beta\chi}^\delta}
	- 4\PD{^\chi}\Dis{^\alpha_\alpha^\beta}\PD{_\delta}\Dis{_{\beta\chi}^\delta}
		\nonumber\\& \hspace{30pt}
	-\frac{2}{3}\PD{_\delta}\Dis{_{\alpha\beta\chi}}\PD{^\delta}\Dis{^{\alpha\beta\chi}}
	\bigg],
\end{align}
is the one found by Fronsdal for spin-three massless propagation~\cite{Fronsdal:1978rb}. 
As far as we are aware, this is the first instance in which this model has been shown to emerge from a symmetry of rank less than two.
A detailed discussion of the implications is presented in a separate work.

\begin{table*}[htbp]
\caption{\label{M1} Models with vector generators. Symmetries are to be applied to~\cref{eq:syms01}, and the model constraints are to be applied to~\cref{VeryFlatLag}. The spectrum is inconsistent (\SampleInconsistent{}), consistent (\SampleConsistent{}), empty (\SampleEmpty{}) or leads to further analysis (\SampleMore{}). Resolved poles may be massive ($\MassiveParticle{J}{P}$), massless ($\MasslessParticle{J}{P}$) or include multiple massless states of unspecified~$J^P$~($\UnknownMasslessParticle{}$).}
\begin{center}
\renewcommand{\arraystretch}{1.7}%
\begin{tabularx}{\linewidth}{c|c|X|c|c}
\hline\hline
	\# & Symmetry & Lagrangian & Unitarity & Spectrum \\
\hline
	\Model{1}{1} &~$\big(t_3 = 0 \big)\wedge\big( t_2 = 0\big)$ &~$\big(\MAGG{4}{_1} = -\tfrac{1}{18}\MAGG{4}{_3} \big)\wedge\big(   \MAGG{4}{_4} = -\tfrac{2}{3} \MAGG{4}{_3} \big)\wedge\big( \MAGG{4}{_7} = -2 \MAGG{4}{_2} + \tfrac{1}{9} \MAGG{4}{_3} \big)\wedge\big(  \MAGG{2}{_2} = -\tfrac{1}{2} \MAGG{2}{_1}\big)$ & \Inconsistent{} &~$\MassiveParticle{0}{+}\wedge\MassiveParticle{1}{-}\wedge\MassiveParticle{2}{+}\wedge\MassiveParticle{3}{-}$ \\
	\Model{1}{2} &~$\big(r_1 = 0 \big)\wedge\big( t_3 = - 2 t_2\big)$ &~$\big(\MAGG{4}{_2} = 0 \big) \wedge \big( \MAGG{4}{_3} =  0 \big) \wedge \big( \MAGG{4}{_4} =  0 \big) \wedge \big( \MAGG{4}{_7} = 0 \big)\wedge\big( \MAGG{2}{_1} =  0 \big) \wedge \big( \MAGG{2}{_2} = 0\big)$ & \Empty{} & --- \\
	\Model{1}{3} &~$\big(r_1 = 0 \big)\wedge\big( t_2 = 0\big)$ &~$\big(\MAGG{4}{_1} = 0 \big)\wedge\big(   \MAGG{4}{_2} = -\tfrac{1}{2}\MAGG{4}{_4} \big)\wedge\big( \MAGG{4}{_3} = -\MAGG{4}{_4} \big)\wedge\big(   \MAGG{4}{_7} = \tfrac{1}{2} \MAGG{4}{_4} \big)\wedge\big( \MAGG{2}{_1} = \MAGG{2}{_2} = 0\big)$ & \Inconsistent{} &~$\UnknownMasslessParticle{}$ \\
	\Model{1}{4} &~$\big(r_1 = 0 \big)\wedge\big( t_3 = -3 t_2\big)$ &~$\big(\MAGG{4}{_1} = -\tfrac{2}{3} \MAGG{4}{_4} \big)\wedge\big(   \MAGG{4}{_2} = -\frac{1}{2} \MAGG{4}{_4} \big)\wedge\big( \MAGG{4}{_3} = 0 \big)\wedge\big(   \MAGG{4}{_7} = -\frac{1}{6} \MAGG{4}{_4} \big)\wedge\big( \MAGG{2}{_1} = \MAGG{2}{_2} = 0\big)$ & \Inconsistent{} &~$\UnknownMasslessParticle{}$ \\
	\Model{1}{5} &~$\big(r_1 = 0 \big)\wedge\big( t_2 = z_2 -z_1 \big)\wedge\big( t_3 = 3 z_1 - 2 z_2\big)$ &~$\big(\MAGG{4}{_1} = \tfrac{1}{3} (\tfrac{z_2}{z_1} + 2 \tfrac{z_1}{z_2} -3 ) \MAGG{4}{_3} \big)\wedge\big( \MAGG{4}{_2} = \tfrac{z_1}{2 z_2} \MAGG{4}{_3} \big)\wedge\big(  \MAGG{4}{_4} = -2 \MAGG{4}{_2} \big)\wedge\big( \MAGG{4}{_7} = \tfrac{z_1 -4 z_2}{6 z_2} \MAGG{4}{_3} \big)\wedge\big( \MAGG{2}{_1} = 0 \big) \wedge \big( \MAGG{2}{_2} = 0\big)$ & \More{} & --- \\
\hline\hline
\end{tabularx}
\end{center}
\end{table*}

\begin{table*}[htbp]
	\caption{\label{m1} Models with vector generators derived as special cases of~\Model{1}{5} in~\cref{M1}. Symmetries are to be applied to~\cref{eq:syms01}, and the model constraints are to be applied to~\cref{VeryFlatLag}. The spectrum is inconsistent (\SampleInconsistent{}), consistent (\SampleConsistent{}), empty (\SampleEmpty{}) or leads to further analysis (\SampleMore{}). Resolved poles may be massive ($\MassiveParticle{J}{P}$), massless ($\MasslessParticle{J}{P}$) or include multiple massless states of unspecified~$J^P$~($\UnknownMasslessParticle{}$).}
\begin{center}
\renewcommand{\arraystretch}{1.7}%
\begin{tabularx}{\linewidth}{c|c|X|c|c}
\hline\hline
	\# & Symmetry & Lagrangian & Unitarity & Spectrum \\
\hline
	\SubModel{1}{2} &~$z_1 = {z_2}/2$ &~$\big(\MAGG{4}{_1} = 0 \big)\wedge\big(   \MAGG{4}{_2} =  \tfrac{1}{4}\MAGG{4}{_3} \big)\wedge\big( \MAGG{4}{_4} =  -\tfrac{1}{2}\MAGG{4}{_3} \big)\wedge\big(   \MAGG{4}{_7} = - \tfrac{7}{12} \MAGG{4}{_3} \big)\wedge\big( \MAGG{2}{_1} =  0 \big) \wedge \big( \MAGG{2}{_2} = 0\big)$ & \Inconsistent{} &~$\UnknownMasslessParticle{}$ \\
	\SubModel{1}{3} &~$z_1 = 2z_2 /3$ &~$\big(\MAGG{4}{_1} = - \tfrac{1}{18} \MAGG{4}{_3} \big)\wedge\big(   \MAGG{4}{_2} =  \tfrac{1}{3}\MAGG{4}{_3} \big)\wedge\big( \MAGG{4}{_4} =  -\tfrac{2}{3} \MAGG{4}{_3} \big)\wedge\big(   \MAGG{4}{_7} = - \tfrac{5}{9} \MAGG{4}{_3} \big)\wedge\big( \MAGG{2}{_1} =  0 \big) \wedge \big( \MAGG{2}{_2} = 0\big)$ & \Inconsistent{} &~$\UnknownMasslessParticle{}$ \\
	\SubModel{1}{4} &~$z_1 = 2 z_2$ &~$\big(\MAGG{4}{_1} = \tfrac{1}{2} \MAGG{4}{_3} \big)\wedge\big(   \MAGG{4}{_2} =  \MAGG{4}{_3} \big)\wedge\big( \MAGG{4}{_4} = -2 \MAGG{4}{_3} \big)\wedge\big(   \MAGG{4}{_7} = -\frac{1}{3} \MAGG{4}{_3} \big)\wedge\big( \MAGG{2}{_1} =  0 \big) \wedge \big(\MAGG{2}{_2} = 0\big)$ & \Consistent{} &~$\MasslessParticle{3}{-}$ \\
	\SubModel{1}{5} &~$z_1 = 4 z_2$ &~$\big(\MAGG{4}{_1} = \tfrac{7}{4} \MAGG{4}{_3} \big)\wedge\big( \MAGG{4}{_2} = -2 \MAGG{4}{_3}  \big)\wedge\big(  \MAGG{4}{_4} = -4 \MAGG{4}{_3} \big)\wedge\big( \MAGG{4}{_7} = 0 \big)\wedge\big( \MAGG{2}{_1} =  0 \big) \wedge \big(\MAGG{2}{_2} = 0\big)$ & \Empty{} & --- \\
\hline\hline
\end{tabularx}
\end{center}
\end{table*}

\begin{figure*}[htbp]
	\includegraphics[width=\linewidth]{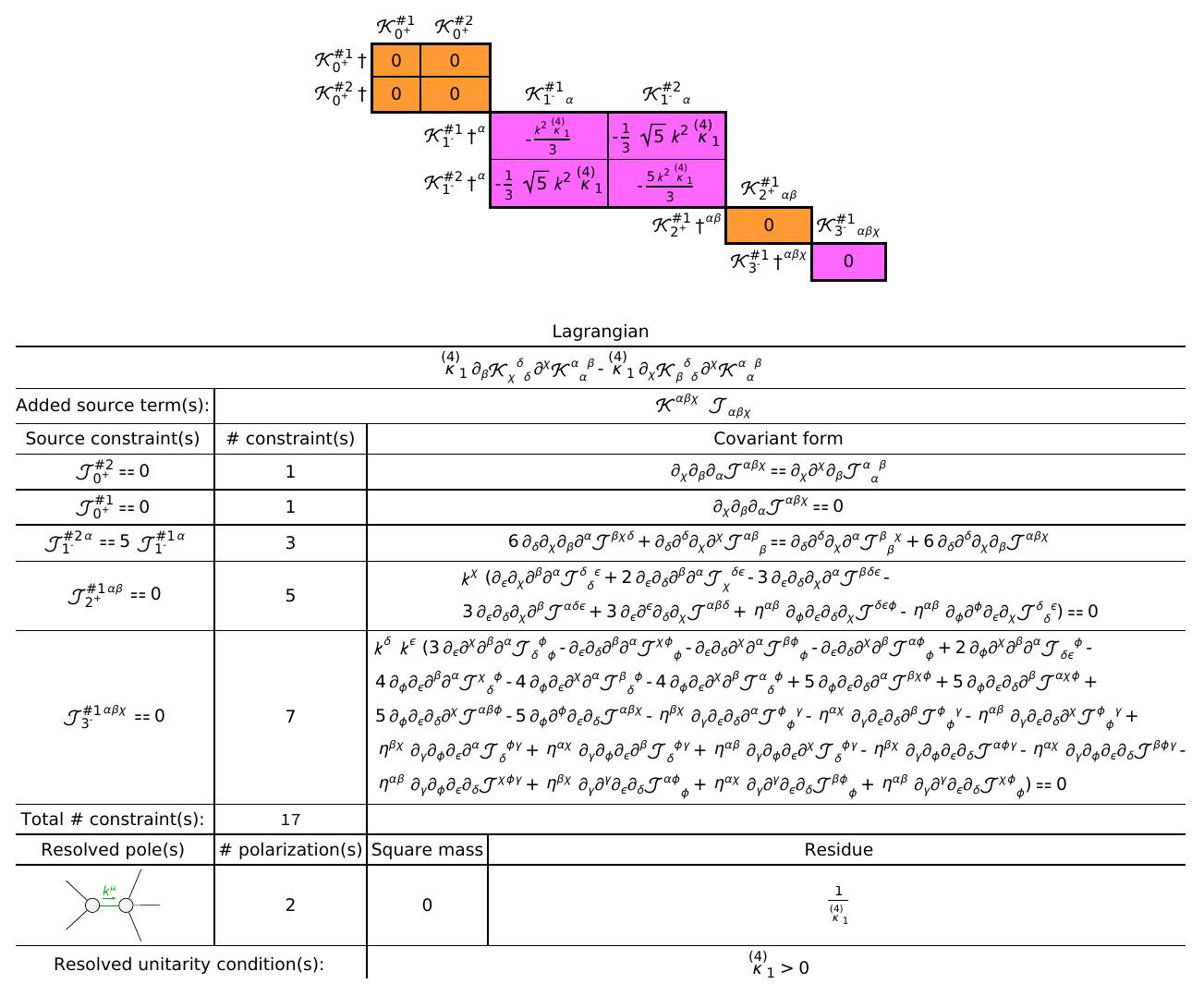}
	\caption{Output generated by \PSALTer{}. The spectrograph of~\Model{2}{3}, as defined in~\cref{M2}. All notation is defined in~\cref{FieldKinematicskNS}.}
\label{ParticleSpectrographM23}
\end{figure*}

\begin{figure*}[htbp]
	\includegraphics[width=\linewidth]{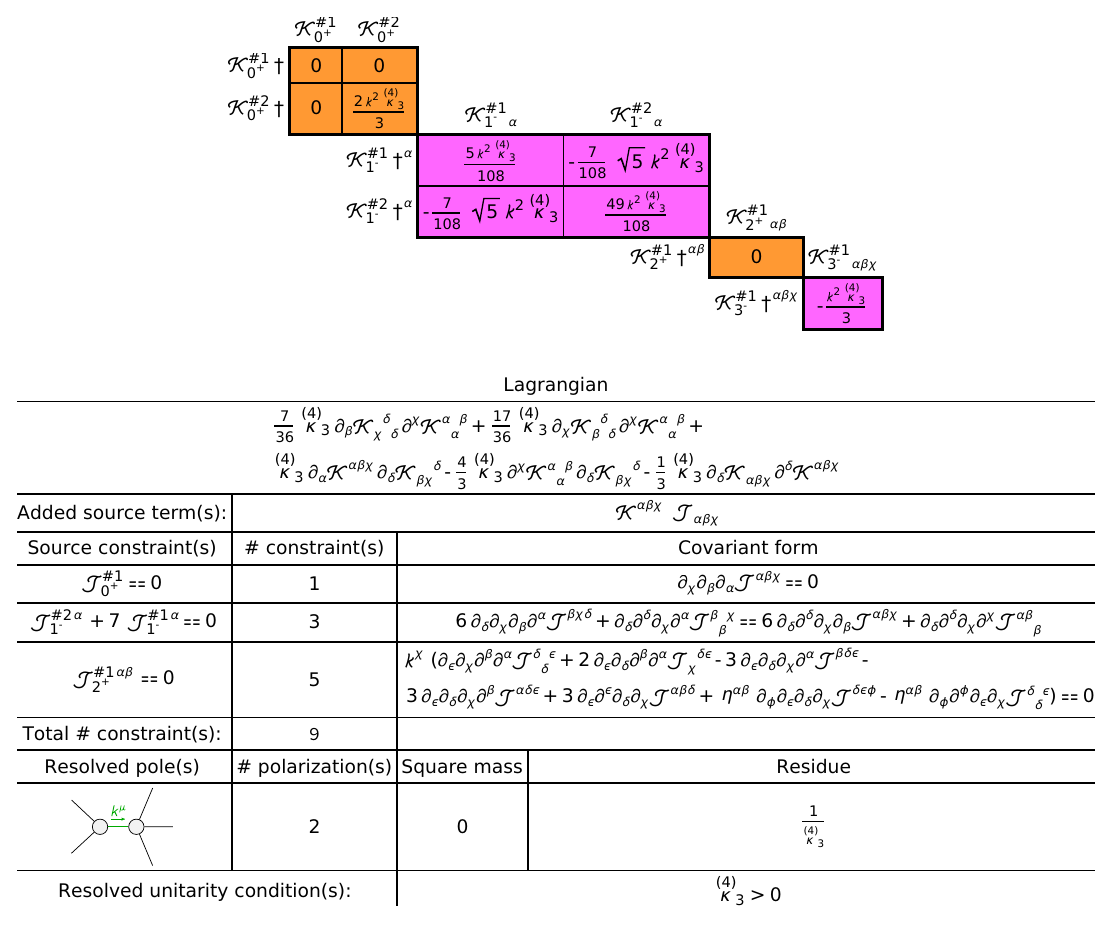}
	\caption{Output generated by \PSALTer{}. The spectrograph of~\SubModel{2}{2}, as defined in~\cref{m2}. All notation is defined in~\cref{FieldKinematicskNS}}
\label{ParticleSpectrographm22}
\end{figure*}

\begin{figure*}[htbp]
	\includegraphics[width=\linewidth]{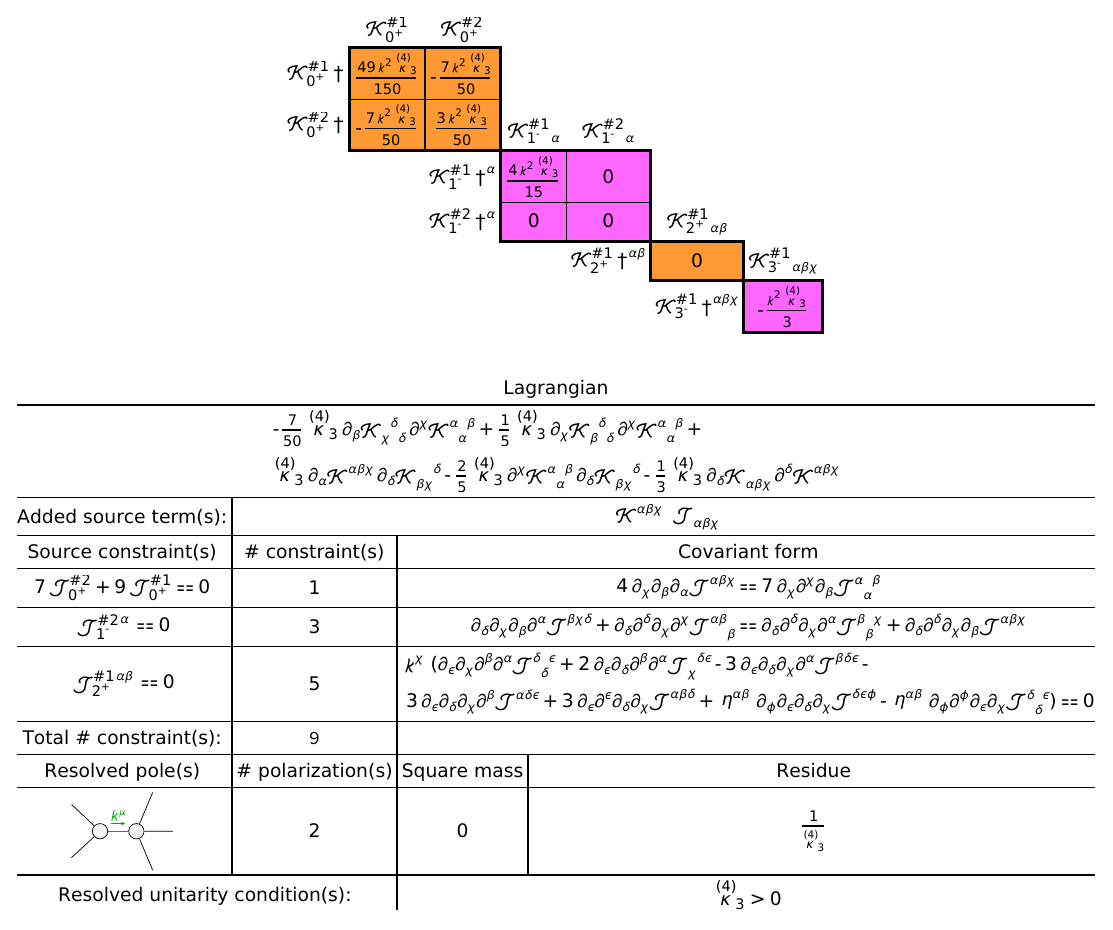}
	\caption{Output generated by \PSALTer{}. The spectrograph of~\SubModel{2}{4}, as defined in~\cref{m2}. All notation is defined in~\cref{FieldKinematicskNS}}
\label{ParticleSpectrographm24}
\end{figure*}

\begin{figure*}[htbp]
	\includegraphics[width=\linewidth]{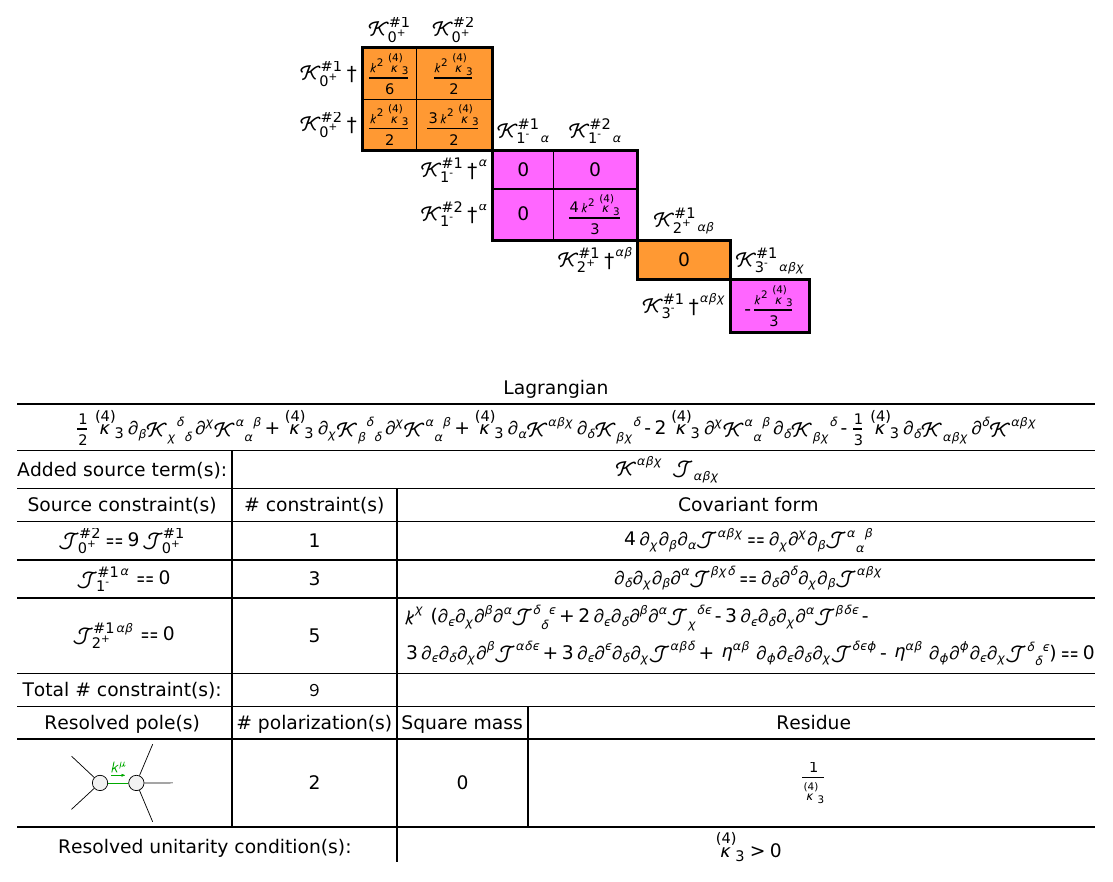}
	\caption{Output generated by \PSALTer{}. The spectrograph of~\SubModel{2}{5}, as defined in~\cref{m2}. All notation is defined in~\cref{FieldKinematicskNS}}
\label{ParticleSpectrographm25}
\end{figure*}

\paragraph*{Tensor generator} The symmetry is given in~\cref{eq:syms02}, and the results of the survey are shown in~\cref{M2}.
Our analysis presents us with four independent solutions.
The first two cases,~\Model{2}{1} and~\Model{2}{2}, are still unable to support consistent model building due to, respectively, multi-particle conflicting propagation and a void spectrum.
Of the latter two cases, the symmetry 
\begin{align}\label{U1flat}
	\tensor{\delta_{(0)}(K)}{_{\alpha\beta\chi}} &=  - \frac{1}{3} \left(\G{_{\chi\beta}} \PD{_{\delta}}\Gen{_{\alpha}^{\delta}} + \G{_{\chi\alpha}} \PD{_{\delta}}\Gen{_{\beta}^{\delta}} + \G{_{\alpha\beta}} \PD{_{\delta}}\Gen{_{\chi}^{\delta}}\right)
	\nonumber \\ & \hspace{15pt}
	+\partial_{a}\xi_{cb} + \partial_{b}\xi_{ca} + \partial_{c}\xi_{ab},
\end{align}
results in the propagation of a healthy massless spin-one particle, as can be seen from the Maxwell-like structure of the corresponding flat Lagrangian
\begin{align}
	\mathcal{L} & = \MAGG{4}{_1} \left(  \PD{_{\beta}}\Dis{_\chi^\delta_\delta} \PD{^{\chi}}\Dis{^\alpha_\alpha^\beta}-\PD{_\chi}\Dis{_\beta^\delta_\delta} \PD{^{\chi}}\Dis{^\alpha_\alpha^\beta}\right),\label{eq:U1flat2}
\end{align}
and the spectrograph in~\cref{ParticleSpectrographM23}.
We reiterate that the notable conclusion of our analysis is not the obvious fact that the Maxwell system can be tuned within the larger quadratic Lagrangian of~$\Dis{_{\mu \nu \rho}}$.
What is remarkable is that such a setup is stable under (symmetry-preserving) radiative corrections because of~\cref{U1flat}, a rank-two gauge symmetry.
Notice also, that had we searched for an Abelian, scalar-generated symmetry for the vector~$\Dis{_{\mu \nu}^{\nu}}$, this would have fallen into our results for scalar generators, which cannot prevent the presence of dangerous operators. 
Using the standard notation in~\cite{Percacci:2020ddy} for the couplings, the curved-space non-linear completion of the Lagrangian in~\cref{eq:U1flat2} --- to the same derivative order as given in~\cref{FlatLag} --- is found to be 
\begin{align}
\mathcal{L} &=
\sqrt{-g}\bigg[\ \frac{1}{2} a_0\, \MAGF{} 
+ \frac{1}{2} c_8\, \MAGF{^{\alpha\beta}_{\alpha}^\chi} \MAGF{_{\beta}^\delta_{\delta\chi}} 
- \frac{1}{2} c_8\, \MAGF{^{\alpha\beta}_{\alpha}^\chi} \MAGF{_{\chi}^\delta_{\delta\beta}} 
	\nonumber\\ & \hspace{15pt}
	- \frac18 (a_0- 4(a_7+a_8))\, \MAGQ{_\alpha^{\chi}_\chi} \MAGQ{^{\alpha\beta}_{\beta}} 
	\nonumber\\ & \hspace{15pt}
	+ \frac{1}{8} (a_0 + 4a_5) \, \MAGQ{_{\alpha\beta\chi}} \MAGQ{^{\alpha\beta\chi}} - \frac{1}{2} a_5 \, \MAGQ{_{\alpha\beta\chi}} \MAGQ{^{\beta\alpha\chi}} 
	\nonumber\\ & \hspace{15pt}
	- \frac12 a_8 \MAGQ{^\alpha_{\alpha}^\beta} \MAGQ{_\beta^\chi_{\chi}}- \frac12 a_7 \MAGQ{^\alpha_{\alpha}^\beta} \MAGQ{^\chi_{\beta\chi}}\bigg] \,,\label{lagrpalspin1}
\end{align}
signalling complete consistency between the non-linear gravitational embedding and the stabilising symmetry.
By contrast, obtaining the non-linear completion of the Fronsdal theory in~\cref{eq:rank1sym,SurpriseFron} poses a significantly more challenging problem. Regarding self-interactions, consistent spin-three theories have only been constructed in anti-de-Sitter space~\cite{Vasiliev:1988sa,Vasiliev:1990en}, while strong obstructions appear to hinder their formulation in Minkowski space~\cite{Bekaert:2010hp}. Interestingly, while consistent spin-three interactions with gravity have been identified~\cite{Boulanger:2006gr, Zinoviev:2008ck}, the non-linear deformation required by the gauge symmetry~\cref{eq:syms4} deviates notably from traditional covariantisation via curved embedding. Consequently, this necessitates the introduction of novel and appropriate invariant methods, distinct from the traditional covariantisation procedure used in~\cref{FlatLag}, i.e. replacing~$\PD{_\mu}$ with~$\tensor{\nabla}{_\mu}$.
Meanwhile, the~\Model{2}{4} case features a parametric symmetry which again demands further dedicated inspection. First, we notice that the provided solution is singular for~$z_1=0$, which suggests that the symmetry may be removed by this particular point in parameter space.
In fact, the~$z_1=0$ case directly reduces to model~\Model{2}{1}. Moreover, the condition~$z_1 = z_2$ reproduces again, up to normalisation, the Fronsdal Lagrangian for a massless spin-three particle~\cite{Fronsdal:1978rb} given already in~\cref{SurpriseFron}. This Lagrangian is known to be supported by a gauge symmetry of the form 
\begin{align} \label{ClassicFron}
	\tensor{\delta_{(0)}(K)}{_{\alpha\beta\chi}} =  \PD{_{\alpha}} \GenTr{_{\beta\chi}} + \PD{_{\beta}} \GenTr{_{\chi\alpha}}  + \PD{_{\chi}} \GenTr{_{\alpha\beta}} \, , 
\end{align}
where~$\GenTr{_{\alpha\beta}}$ is a \emph{traceless} rank-two generator.
It is readily confirmed that our particular tuning among the~$\big\{r_i\big\}$ parameters results in the trace being removed from the generator~$\Gen{_{\mu \nu}}$. 
Given this information, and that generic values of the free parameters $z_i$ continuously connect to the Fronsdal Lagrangian, it is natural to ask whether model~\Model{2}{4} offers a more general description of spin-three massless propagation. This is \emph{not} the case. We can explicitly prove this point by observing that the simple field redefinition
\begin{align}
	\Dis{_{\alpha \beta \gamma}} & \mapsto f_1 \Dis{_{\alpha \beta \gamma}} \nonumber\\
	& \hspace{10pt}+ f_2 \left(\G{_{\sigma \alpha}} \G{_{\beta \gamma}} + \G{_{\sigma \beta}} \G{_{\alpha \gamma}} + \G{_{\sigma \gamma}} \G{_{\alpha \beta}} \right) \Dis{^{\sigma \mu}_{\mu}} \, , 
\end{align}
when applied to~\cref{SurpriseFron}, reproduces our parametric model with $f_1 = z_1/\sqrt{2}$ and $f_2 = (z_1-z_2)/6\sqrt{2}$. 
Consistent with this observation, we can apply the same techniques used to obtain~\cref{m1} from~\cref{M1} to the parametric model~\Model{2}{4} to produce the seemingly special branches shown in~\cref{m2}. These all propagate the spin-three massless particle, shown for example in~\cref{ParticleSpectrographm22,ParticleSpectrographm24,ParticleSpectrographm25}.
We can therefore conclude rank-two symmetries stabilise the healthy propagation of spin-one and spin-three particles.

\begin{table*}[htbp]
\caption{\label{M2} Models with tensor generators. Symmetries are to be applied to~\cref{eq:syms02}, and the model constraints are to be applied to~\cref{VeryFlatLag}. The spectrum is inconsistent (\SampleInconsistent{}), consistent (\SampleConsistent{}), empty (\SampleEmpty{}) or leads to further analysis (\SampleMore{}). Resolved poles may be massive ($\MassiveParticle{J}{P}$), massless ($\MasslessParticle{J}{P}$) or include multiple massless states of unspecified~$J^P$~($\UnknownMasslessParticle{}$).}
\begin{center}
\renewcommand{\arraystretch}{1.7}%
\begin{tabularx}{\linewidth}{c|c|X|c|c}
\hline\hline
	\# & Symmetry & Lagrangian & Unitarity & Spectrum \\
\hline
	\Model{2}{1} &~$\big( r_1 = -\frac{1}{3} r_3 \big)\wedge\big( r_2 = -\frac{1}{6} r_3\big)$ &~$\big(\MAGG{4}{_3} =  0 \big) \wedge \big( \MAGG{4}{_4} =  0 \big) \wedge \big( \MAGG{4}{_7} = 0 \big)\wedge\big( \MAGG{2}{_1} = 0\big)$ & \Inconsistent{} &~$\MassiveParticle{0}{+}\wedge\MassiveParticle{1}{-}$ \\
	\Model{2}{2} &~$\big(r_1 = -r_3 \big)\wedge\big( r_2 = 0\big)$ &~$\big(\MAGG{4}{_2} = - \MAGG{4}{_1} \big)\wedge\big(   \MAGG{4}{_3} = -4 \MAGG{4}{_1} \big)\wedge\big( \MAGG{4}{_4} = 0 \big)\wedge\big(   \MAGG{4}{_7} = \frac{4}{3} \MAGG{4}{_1} \big)\wedge\big( \MAGG{2}{_1} =  0 \big) \wedge \big( \MAGG{2}{_2} = 0\big)$ & \Empty{} & --- \\
	\Model{2}{3} &~$\big(r_1 = -\frac{1}{3} r_3 \big)\wedge\big( r_2 = 0\big)$ &~$\big(\MAGG{4}{_2} = -\MAGG{4}{_1} \big)\wedge\big( \MAGG{4}{_3} =  0 \big) \wedge \big(\MAGG{4}{_4} =  0 \big) \wedge \big( \MAGG{4}{_7} = 0 \big)\wedge\big( \MAGG{2}{_1} =  0 \big) \wedge \big( \MAGG{2}{_2} = 0\big)$ &~\Consistent{} &~$\MasslessParticle{1}{-}$ \\
	\Model{2}{4} &~$r_1 = \tfrac{z_1-z_2}{3}$ &~$\big(\MAGG{4}{_1}  = -\tfrac{\left(z_1^2 - 8 z_1 z_2 - 2 z_2^2\right)}{18 z_1^2}\MAGG{4}{_3}  \big)\wedge\big( \MAGG{4}{_2} = \tfrac{\left(2 z_1^2 + 2 z_1 z_2 + 5 z_2^2\right)}{9 z_1^2}\MAGG{4}{_3} \big)\wedge\big( \MAGG{4}{_4} = -\tfrac{2\left(z_1 + 2 z_2\right)}{3 z_1} \MAGG{4}{_3} \big)\wedge\big( \MAGG{4}{_7} = -\tfrac{1}{3} \MAGG{4}{_3} \big)\wedge\big( \MAGG{2}{_1} =0 \big)\wedge\big(  \MAGG{2}{_2} = 0\big)$ &~\More{} & --- \\
\hline\hline
\end{tabularx}
\end{center}
\end{table*}

\begin{table*}[htbp]
\caption{\label{m2} Models with tensor generators derived as special cases of~\Model{2}{4} in~\cref{M2}. Symmetries are to be applied to~\cref{eq:syms01}, and the model constraints are to be applied to~\cref{VeryFlatLag}. The spectrum is inconsistent (\SampleInconsistent{}), consistent (\SampleConsistent{}), empty (\SampleEmpty{}) or leads to further analysis (\SampleMore{}). Resolved poles may be massive ($\MassiveParticle{J}{P}$), massless ($\MasslessParticle{J}{P}$) or include multiple massless states of unspecified~$J^P$~($\UnknownMasslessParticle{}$).}
\begin{center}
\renewcommand{\arraystretch}{1.7}%
\begin{tabularx}{\linewidth}{c|c|X|c|c}
\hline\hline
	\# & Symmetry & Lagrangian & Unitarity & Spectrum \\
\hline
	\SubModel{2}{2} &~$z_1=2z_2$ &~$\big(\MAGG{4}{_1}=\tfrac{7}{36}\MAGG{4}{_3} \big)\wedge\big( \MAGG{4}{_2}=\tfrac{17}{36}\MAGG{4}{_3} \big)\wedge\big( \MAGG{4}{_4}=-\tfrac{4}{3}\MAGG{4}{_3} \big)\wedge\big( \MAGG{4}{_7}=-\tfrac{1}{3}\MAGG{4}{_3} \big)\wedge\big( \MAGG{2}{_1}=0 \big)\wedge\big( \MAGG{2}{_2}=0\big)$ & \Consistent{} &~$\MasslessParticle{3}{-}$ \\
	\SubModel{2}{3} &~$z_1=-2z_2$ &~$\big(\MAGG{4}{_1}=-\tfrac{1}{4}\MAGG{4}{_3} \big)\wedge\big( \MAGG{4}{_2}=\tfrac{1}{4}\MAGG{4}{_3} \big)\wedge\big( \MAGG{4}{_4}=0 \big)\wedge\big( \MAGG{4}{_7}=-\tfrac{1}{3}\MAGG{4}{_3} \big)\wedge\big( \MAGG{2}{_1}=0 \big)\wedge\big( \MAGG{2}{_2}=0\big)$ & \Consistent{} &~$\MasslessParticle{3}{-}$ \\
	\SubModel{2}{4} &~$z_1=-5z_2$ &~$\big(\MAGG{4}{_1}=-\tfrac{7}{50}\MAGG{4}{_3} \big)\wedge\big( \MAGG{4}{_2}=\tfrac{1}{5}\MAGG{4}{_3} \big)\wedge\big( \MAGG{4}{_4}=-\tfrac{2}{5}\MAGG{4}{_3} \big)\wedge\big( \MAGG{4}{_7}=-\tfrac{1}{3}\MAGG{4}{_3} \big)\wedge\big( \MAGG{2}{_1}=0 \big)\wedge\big( \MAGG{2}{_2}=0\big)$ & \Consistent{} &~$\MasslessParticle{3}{-}$ \\
	\SubModel{2}{5} &~$z_1=z_2$ &~$\big(\MAGG{4}{_1}=\tfrac{1}{2}\MAGG{4}{_3} \big)\wedge\big( \MAGG{4}{_2}=\MAGG{4}{_3} \big)\wedge\big( \MAGG{4}{_4}=-2\MAGG{4}{_3} \big)\wedge\big( \MAGG{4}{_7}=-\tfrac{1}{3}\MAGG{4}{_3} \big)\wedge\big( \MAGG{2}{_1}=0 \big)\wedge\big( \MAGG{2}{_2}=0\big)$ & \Consistent{} &~$\MasslessParticle{3}{-}$ \\
\hline\hline
\end{tabularx}
\end{center}
\end{table*}

\section{Conclusions}\label{Sec:Concl}

\paragraph*{Results} We have found that parity-preserving metric-affine gravity with a symmetric distortion tensor is able to support the healthy propagation of massless spin-one and spin-three particles due to symmetries that protect against radiative instabilities. In the spin-three case, this assumes that a non-linear completion of the theory exists. The methods used do not rely on the arbitrary tuning of couplings to achieve a healthy spectrum, and do not result in accidental symmetries.

\paragraph*{Need for masses} Our methods have led exclusively to theories propagating massless modes. New light degrees of freedom are undesirable, since there is no evidence for them in the low-energy phenomena. This should not, however, disqualify such models as candidate foundations for metric-affine theories. Indeed, the point is precisely that these models are supposed to be foundational, and there are at least two ways in which finite masses may be acquired through further responsible model-building procedures:
\begin{description}
	\item [Symmetry-breaking] By coupling each model to a compensator field in such a way that the motivating gauge symmetry is preserved. Masses may then be acquired through Stuekelberg-like condensation of the compensator. This is the effect that gives rise to the massses of the~$\text{W}^{\pm}$ and~$Z^0$ bosons of the weak sector, through the Higgs mechanism.
	\item [Non-linear completion] The symmetry may entail interactions whose non-perturbative quantum effects radically alter the low-energy spectrum. For example, the spectrum of the strong sector is dominated by the lightest hadrons rather than the massless gluons.
\end{description}
Both of these mechanisms have precedent in the real physics of the standard model. Note how, in either case, it was instructive to first establish the underlying~$\mathrm{SU}(2)$- and~$\mathrm{SU}(3)$-symmetric models before worrying about mass acquisition. Following~\cite{Donoghue:2016vck}, we advocate for similar approaches in metric-affine gravity.\footnote{Note that this approach calls for additional careful work. It contrasts strongly with the usual practice in non-Riemannian gravity of tuning the couplings so as to populate only a unitary massive spectrum. Whilst this latter method is superficially consistent with constraints on light new species, it is basically arbitrary for the reasons set out in~\cref{Sec:Pred}, and moreover it does not actually lead to any predictions about the values of the masses.}

\paragraph*{Further work} The approach should next be applied to full metric-affine gravity, without the somewhat artificial restrictions imposed here, of parity invariance and a totally symmetric distortion tensor. The methods can also be extended to admit un-free symmetries. Apart from such systematic surveys, it is worth considering whether previously proposed models of metric-affine gravity qualify as good foundations for EFTs, despite their not being derived using the methods of this paper. Chief among such models is metric-affine gravity which is invariant under the so-called \emph{extended projective} (EP) symmetry~\cite{Barker:2024dhb}. EP invariance is predictive, in that it leads to a uniquely healthy massive pseudoscalar particle, but this is only true when starting from the scalar invariants of~$\MAGF{_{\mu\nu}^\rho_\sigma}$,~$\MAGT{_\mu^\alpha_\nu}$ and~$\MAGQ{_{\lambda\mu\nu}}$ as an initial operator basis --- contrary to our approach in~\cref{VeryFlatLag}. Thus, EP invariance blends aspects of EFT and the geometry in~\cref{NonRiemannianSchematic}. Quantum corrections will be crucial in determining the physical utility of EP invariance.

\begin{acknowledgments}
This work was improved by useful discussions with Dario Francia, Dražen Glavan, Mike Hobson, Giorgos Karananas, Anthony Lasenby, Roberto Percacci, Syksy R\"as\"anen and Sebastian Zell.

This work used the DiRAC Data Intensive service~(CSD3 \href{www.csd3.cam.ac.uk}{www.csd3.cam.ac.uk}) at the University of Cambridge, managed by the University of Cambridge University Information Services on behalf of the STFC DiRAC HPC Facility~(\href{www.dirac.ac.uk}{www.dirac.ac.uk}). The DiRAC component of CSD3 at Cambridge was funded by BEIS, UKRI and STFC capital funding and STFC operations grants. DiRAC is part of the UKRI Digital Research Infrastructure.

This work also used the Newton compute server, access to which was provisioned by Will Handley using an ERC grant.

W.~B. is grateful for the support of Girton College, Cambridge, Marie Skłodowska-Curie Actions and the Institute of Physics of the Czech Academy of Sciences. The work of C.~M. was supported by the Estonian Research Council grant PRG1677. A.~S. acknowledges financial support from the ANID CONICYT-PFCHA/DoctoradoNacional/2020-21201387.

Co-funded by the European Union (Physics for Future – Grant Agreement No. 101081515). Views and opinions expressed are however those of the author(s) only and do not necessarily reflect those of the European Union or European Research Executive Agency. Neither the European Union nor the granting authority can be held responsible for them.
\end{acknowledgments}

\bibliography{Manuscript}

\appendix

\end{document}